\documentclass[10pt]{article}
\usepackage{jhep-mod}
\usepackage{bm}
\usepackage{amssymb,amsmath,amsthm}
\usepackage{mathrsfs}

\usepackage[utf8]{inputenc}
\usepackage{enumerate}
\usepackage{appendix}
\usepackage{graphicx}
\usepackage{dcolumn}
\usepackage{bm}
\usepackage{multirow}
\usepackage{float}
\usepackage{cuted}

\newcommand{\upstar}[1]{\overset{\star}{#1}{}}
\newcommand{\der}[2]{\overline{\nabla}_{#2}}

\begin{document}
	
	\title{Stability in quadratic torsion theories}
	
	\author{Teodor Borislavov Vasilev, Jose A. R. Cembranos, Jorge Gigante Valcarcel {\rm and} Prado Martín-Moruno}
	\affiliation{Departamento de Física Teórica I, Universidad Complutense de Madrid, E-28040 Madrid, Spain}
	\emailAdd{teodorbo@ucm.es}
	\emailAdd{cembra@fis.ucm.es}
	\emailAdd{jorgegigante@ucm.es}
	\emailAdd{pradomm@ucm.es}
	
	\abstract{
		We revisit the definition and some of the characteristics of quadratic theories of gravity with torsion. 
		We start from a Lagrangian density quadratic in the curvature and torsion tensors.
		By assuming that General Relativity should be recovered when torsion vanishes and investigating the behaviour of the vector and pseudovector torsion fields in the weak-gravity regime,
		we present a set of \emph{necessary} conditions for the stability of these theories.
		Moreover, we explicitly obtain the gravitational field equations using the Palatini variational principle with the metricity condition implemented via a Lagrange multiplier.
	}
	
	\maketitle

	\section{Introduction}
	\label{intro}
	General Relativity (GR) radically changed our understanding about the Universe. The predictions of this elegant theory have been confirmed up to the date \cite{expGR,Abbott:2016blz}.
	In order to fit extragalactic and cosmological observational data, however, the presence of a non-vanishing cosmological constant and six times more dark matter than ordinary one have
	to be assumed in this framework \cite{eleft}. In addition, the observed value of this cosmological constant differs greatly from the value expected for the vacuum energy.
	On the other hand, while the strong and electroweak forces are renormalizable gauge theories, that is not the case for GR, and
	the compatibility of GR with the quantum realm is still a matter of debate.
	Given this situation, there has been a renewed interest in alternative theories of gravity, which modify the predictions of GR.
	
	A particular approach to formulate alternative theories of gravity involves an extension of the geometrical treatment that covers the microscopic properties of matter \cite{torsionlibro}. 
	It should be noted that the mass is not enough to characterize  particles at the quantum level given that they have an other independent \textit{label}, that is, the spin. 
	Whereas at macroscopic scales the energy-momentum tensor is enough to describe the source of gravity, a description of the space-time distribution of the spin density is needed at microscopic scales. 
	Moreover, there are macroscopic configurations that may also need a description of the spin distribution, as super-massive objects (e.~g.~black holes or neutron stars with nuclear polarization).
	In this spirit, a new geometrical concept should be related to the spin distribution in the same way that space-time curvature is related to the energy-momentum distribution. 
	Torsion is a natural candidate for this purpose \cite{torsionlibro,yao}. Furthermore, an important advantage of a theory of gravity with torsion is that it can be 
	formulated as a gauge theory \cite{New,Class,Obukhov}.

	Since 1924 many authors have considered theories of gravity in a Riemann--Cartan $U_4$ space-time. 
	In this manifold the non-vanishing torsion can be coupled to the intrinsic spin density of matter and, in this way, the spin part of the Poincar\'e group can change the geometry of the manifold as the 
	energy-momentum tensor does it. 
	The first attempt to introduce torsion in a theory of gravity was the Einstein-Cartan theory, which is a reformulation of GR in a $U_4$ space-time. In this theory
	the scalar curvature of the Einstein--Hilbert action is constructed from a $U_4$ connection instead of using the Christoffel symbols.  However, the resulting theory was not completely satisfactory
	because the field equations relate the torsion and its source in an algebraic way and, therefore, torsion is not dynamical. 
	Hence the torsion field vanishes in vacuum and the Einstein-Cartan theory collapses to GR except for unobservable corrections to the energy-momentum tensor \cite{torsionlibro}.
	In order to obtain a theory with propagating torsion, we need to consider an action that is at least quadratic in the curvature tensors \cite{torsionlibro,New,Class,Obukhov,Neville1,Birkov,Flatness}. Moreover, an important 
	advantage of adding quadratic terms ${\cal{R}}^2$ to the Einstein--Hilbert action is the possibility of 
	making the theory renormalizable \cite{Neville1}. In addition,  it can be shown \cite{torsionlibro,New} that, considering a gauge description, the torsion and curvature tensors
	correspond to the field-strength tensors of the gauge potentials of the Poincar\'e group $(e_{\mu}^{\ a }, w_{\mu}^{\ a b })$, which are the vierbein and the local Lorentz connection, respectively. 
	Thus, a pure ${\cal{R}}^2$ gauge theory of gravity has some resemblance to electro-weak and strong theories.
	
	From an experimental point of view there have been many attempts to detect torsion or to set an upper bound to its gravitational effects. 
	One of the most debated attempts was the use of the Gravity Probe B experiment to measure torsion effects \cite{GPB}. Nevertheless, this experiment was criticized because torsion will never coupled to the gyroscopes installed in the satellite \cite{Hehl:2013qga}. 
	Therefore, this probe cannot measure the  gravitational effects due to torsion. On the other hand, other unsuccessful experiments aimed to constrain torsion with accurate measurements on the perihelion advance and 
	the orbital geodetic effect of a satellite \cite{moon}. 
	The experimental difficulty is due to the need of dealing with elementary particles with spin to obtain a maximal coupling with torsion. 
	
	In this paper we present a self-contained introduction to quadratic theories of gravity with torsion in the geometrical approach (gauge treatment is not considered). 
	We partly recover known results about the stability of these theories using simple methods.
	Thus, we simplify the existent mathematical treatment and reinforce the critical discussion about some controversial results published in the literature.
	
	The paper is organized as follows: In Sec.~\ref{sec:basics}, we present a general introduction to the basic concepts on general affine geometries and introduce the conventions used throughout the paper. 
	In Sec.~\ref{sec:Analisis}, we present our main results. In the first place, we consider a Lagrangian density  quadratic in the curvature and torsion tensors.
	In Sec.~\ref{sec:ecuaciones}, we discuss the different methods presented in the literature to obtain the field equations and explicitly derive them in the Palatini formalism.
	In Sec.~\ref{sec:reduccionGR}, we obtain conditions on the parameters of the Lagrangian necessary to avoid large deviations from GR and instabilities. 
	Then, in Sec.~\ref{sec:stability}, we analyse the Lagrangian density with the aim of setting necessary conditions for avoiding ghost and tachyon instabilities.
	The conclusions are summarized in Sec.~\ref{sec:conclus}. We relegate some calculations and further comments to the appendices:
	in Appendix \ref{G-B} we include the Gauss--Bonnet term in Riemann--Cartan geometries; in Appendix \ref{variations} we include detailed expressions necessary to obtain the equations of the dynamics using
	the Palatini formalism; in Appendix \ref{Ap:Sources} we discuss the source terms of these equations; and, in Appendix \ref{QuadraticTerms}, we include relevant expressions for the study of the 
	vector and pseud-vector torsion fields around Minkowski.


	\section{Basic concepts and conventions \label{sec:basics}}
	
	The geometric structure of a manifold can be catalogued by the properties of the affine connection. A general affine connection $\tilde{\Gamma}$ provides three main characteristics: curvature, torsion, 
	and non-metricity. Combinations of these quantities in the affine connection generate the geometric structure \cite{yao}. In GR it is assumed that the space-time geometry is described by a Riemannian 
	manifold, thus the affine connection reduces to the so called Levi-Civita connection and the gravitational effects are only produced by the consequent curvature in terms of the metric tensor alone. 
	Nevertheless, in a general geometrical theory of gravity the gravitational effects are generated by the whole connection, which involves a post-Riemannian approach described by curvature, torsion 
	and non-metricity.
	In this scheme, there are many ways to deal with torsion and non-metricity due to different conventions. For that reason, it is important to set the conventions and definitions used throughout this work. 
	Thus, the notation assumed for the symmetric and the antisymmetric part of a tensor $A$ are
	\begin{align}
		& A_{(\mu_1 \cdots \mu_s)}\equiv\frac{1}{s!}\sum_{\pi\in P(s)} A_{\pi (\mu_1) \cdots \pi (\mu_s )}  \, ,\\
		& A_{[\mu_1 \cdots \mu_s]}\equiv\frac{1}{s!}\sum_{\pi\in P(s)} \textup{sgn}(\pi)A_{\pi (\mu_1) \cdots \pi (\mu_s )} \, ,
	\end{align}
	respectively, where $P(s)$ is the set of all the permutations of $1, ..., s$ and $\textup{sgn}(\pi)$ is positive for even permutations whereas it is negative for odd permutations.
	
	In the first place, the Cartan torsion is defined as the antisymmetric part of the affine connection as \cite{torsionlibro,VarPrinGR,f(R)cosmology with torsion,f(R)}
	\begin{equation}
		T^{\mu}_{\cdot \nu \sigma}\equiv \tilde{\Gamma}^{\mu}_{\cdot [ \nu \sigma ] } .
		\label{defTorsion}
	\end{equation}
	Note that a dot $``."$ appears below the index $\mu$ indicating the position that it takes when is lowered with the metric. 
	As the difference of two connections transforms as a tensor, then the Cartan torsion is a tensor. Thus, 
	from now on we call it just torsion and emphasize that it cannot be eliminated with a suitable change of coordinates.

	In the second place, non-metricity can also be described by a third rank tensor. This is
	\begin{equation}
		Q_{\rho\mu\nu}\equiv \tilde{\nabla}_{\rho}g_{\mu\nu} \, ,
		\label{defQ}
	\end{equation}
	where $\tilde{\nabla}$ is the covariant derivative defined from the affine connection $\tilde{\Gamma}$. The non-metricity tensor is usually split into a trace vector $\omega_{\rho}\equiv \frac{1}{4}Q^{\ \ \ \nu}_{\rho\nu\cdot}$, called the Weyl vector \cite{nonMetricity}, and a traceless part $\overline{Q}_{\rho\mu\nu}$,
	\begin{equation}
		Q_{\rho\mu\nu}=w_{\rho}g_{\mu\nu}+\overline{Q}_{\rho\mu\nu}.
	\end{equation}
	It should be noted that there are manifolds with non-metricity where the cancellation of the $\omega_{\rho}$ or the traceless part of $Q$ are demanded.

	Since the general connection  $\tilde{\Gamma}$ is asymmetric in the last two indices, a convention is needed  for the covariant derivative of a tensor. Let be $A^{\mu_1 \cdots \mu_r}_{\cdot \ \ \cdots \ \cdot \ \nu_1 \cdots \nu_s}$  the components of a tensor type $(r, s)$, then
	\begin{eqnarray}
		&\tilde{\nabla}_{\rho}& A^{\mu_1 \cdots \mu_r}_{\cdot \ \ \cdots \ \cdot \ \nu_1 \cdots \nu_s} \equiv  \partial_{\rho}A^{\mu_1 \cdots \mu_r}_{\cdot \ \ \cdots \ \cdot \ \nu_1 \cdots \nu_s} \nonumber\\
		&+&\sum_{i=1}^{r}\tilde{\Gamma}^{\mu_i}_{\cdot \lambda \rho}A^{\mu_1 \cdot \lambda\cdot \mu_r}_{\cdot \ \ \cdots \ \cdot \ \nu_1 \cdots \nu_s} 
		-\sum_{j=1}^{s}\tilde{\Gamma}^{\lambda}_{\cdot \nu_j \rho}A^{\mu_1 \cdots \mu_r}_{\cdot \ \ \cdots \ \cdot \ \nu_1 \cdot\lambda\cdot \nu_s} .
	\end{eqnarray}
	It is important to emphasize the syntax of the lower indices in the affine connections, this is the index $\rho$ of the derivative is written in the last position in the affine connection. 
	
	Using the definitions presented in this section, the general connection  $\tilde{\Gamma}$ is written as \cite{torsionlibro,VarPrinGR,shapiro_torsion}
	\begin{equation}
		\tilde{\Gamma}^{\mu}_{\cdot \nu \sigma}= \Gamma^{\mu}_{\cdot\nu \sigma}+ W^{\mu}_{\cdot \nu \sigma} \, ,
		\label{connectiontilde}
	\end{equation}
	with $\Gamma^{\mu}_{\cdot\nu \sigma}$ the Levi-Civita connection
	\begin{equation}
		\Gamma^{\mu}_{\cdot\nu \sigma}=\frac{1}{2}g^{\mu \rho}\Delta_{\sigma\nu\rho}^{\alpha\beta\gamma}\partial_{\alpha}g_{\beta \gamma} \, ,
		\label{levicivita}
	\end{equation}
	which is expressed in a compact form by the permutation tensor \cite{10}
	\begin{equation}
		\Delta_{\sigma\nu\rho}^{\alpha\beta\gamma}=\delta_{\sigma}^{\ \alpha}\delta_{\nu}^{\ \beta}\delta_{\rho}^{\ \gamma}+\delta_{\nu}^{\ \alpha}\delta_{\rho}^{\ \beta}\delta_{\sigma}^{\ \gamma}-\delta_{\rho}^{\ \alpha}\delta_{\sigma}^{\ \beta}\delta_{\nu}^{\ \gamma}\, ,
		\label{DeltaPermutation}
	\end{equation}
	and the additional tensor $W^{\mu}_{{}^{.} \nu \sigma}$ defined by the following expression:
	\begin{equation}
		W^{\mu}_{\cdot \nu \sigma}= K^{\mu}_{{}^{.} \nu \sigma}+\frac{1}{2}\left( Q^{\mu}_{\cdot\nu\sigma}-Q^{\ \mu}_{\sigma\cdot\nu}-Q^{\ \mu}_{\nu\cdot\sigma}\right)\, ,
	\end{equation}
	where $K^{\mu}_{{}^{.} \nu \sigma}$ is called the contortion tensor,
	\begin{equation}
		K^{\mu}_{{}^{.} \nu \sigma}=T^{\mu}_{{}^{.} \nu \sigma}-T^{ \ \mu}_{\nu {}^{.} \sigma}-T^{ \ \mu}_{\sigma {}^{.} \nu} \ .
		\label{eq:cotorsion}
	\end{equation}
	Note that $Q_{\rho\mu\nu}$ is symmetric in the last two indices while $T^{\mu}_{\cdot \nu \sigma}$ is antisymmetric in these indices. However, contortion, $K^{\mu}_{{}^{.} \nu \sigma}$, is antisymmetric in the first pair of indices. This property ensures the existence of a metric-compatible connection when the non-metricity tensor vanishes.
	
	Furthermore, it is useful to write torsion through its three irreducible components. These are \cite{shapiro_torsion}
	\begin{itemize}
		\item[\textit{i)}] the trace vector $ T^{\mu}_{{}^{.} \nu \mu}\equiv T_{\nu}$.
		\item[\textit{ii)}] the pseudo-trace axial vector $S^{\nu}\equiv\epsilon^{\alpha \beta \sigma \nu}T_{\alpha \beta \sigma}$.
		\item[\textit{iii)}] the tensor $q^{\alpha}_{{}^{.} \beta \sigma}$ , which satisfies  $ q^{\alpha}_{{}^{.} \beta \alpha}=0 $ and \newline $ \epsilon^{\alpha \beta \sigma \nu} q_{\alpha \beta \sigma}=0 $.
	\end{itemize}
	Thus, the torsion field can be rewritten as
	\begin{equation}
		T^{\alpha}_{\cdot \beta \mu}=\frac{1}{3}(T_{\beta} \delta^{\alpha}_{\ \mu}- T_{\mu} \delta^{\alpha}_{\  \beta})+ \frac{1}{6}g^{\alpha\sigma}\epsilon_{\sigma \beta \mu \nu} S^{\nu} + q^{\alpha}_{\cdot \beta \mu  }  \ .
		\label{eq:torsionDESCOMP.}
	\end{equation}
	
	The introduction of these new geometrical degrees of freedom leads to the generalization of the usual definition of the curvature tensor in the Riemann space-time, $[\nabla_{\rho},\nabla_{\sigma}]V^{\mu}=R^{\mu}_{ \cdot \nu \rho  \sigma}V^{\nu} $, by the following commutative relations associated with a connection $\tilde{\Gamma}$:
	\begin{equation}
		[\tilde{\nabla}_{\rho},\tilde{\nabla}_{\sigma}]V^{\mu}= \tilde{R}^{\mu}_{\cdot \nu \rho  \sigma}V^{\nu}+2T^{\alpha}_{\cdot \rho \sigma}\tilde{ \nabla}_{\alpha}V^{\mu}\, ,
		\label{eq:IdentRicci}
	\end{equation}
	where the curvature tensor reads
	\begin{equation}
		\tilde{R}^{\mu}_{\cdot \nu \rho  \sigma}= \partial_{\rho}\tilde{\Gamma}^{\mu}_{\cdot \nu \sigma} - \partial_{\sigma}\tilde{\Gamma}^{\mu}_{\cdot \nu \rho} + \tilde{\Gamma}^{\mu}_{\cdot \lambda \rho}\tilde{\Gamma}^{\lambda}_{\cdot \nu \sigma} -  \tilde{\Gamma}^{\mu}_{\cdot \lambda \sigma}\tilde{\Gamma}^{\lambda}_{\cdot \nu \rho} \ .
		\label{eq:RiemannTilde}
	\end{equation}
	Using Eq.~(\ref{connectiontilde}), the curvature tensor can be rewritten as
	\begin{eqnarray}
		\tilde{R}^{\mu}_{\cdot \nu \rho  \sigma}&=&R^{\mu}_{ \cdot  \nu \rho  \sigma}+ \nabla_{\rho} W^{\mu}_{\cdot\nu \sigma} - \nabla_{\sigma}W^{\mu}_{\cdot \nu \rho} + W^{\mu}_{\cdot \lambda \rho}W^{\lambda}_{\cdot \nu \sigma}-  W^{\mu}_{\cdot \lambda \sigma}W^{\lambda}_{\cdot \nu \rho} ,
		\label{RiemannTilde2}
	\end{eqnarray}
	with $R^{\mu}_{ \cdot  \nu \rho  \sigma}$ the curvature tensor of the Riemann space-time, commonly called Riemann tensor, and $\nabla$ the covariant derivative constructed from the Levi-Civita connection.
	
	On the other hand, the generalization of the two Bianchi identities can be computed from the expression (\ref{eq:RiemannTilde}). Taking into account Eq.~(\ref{defTorsion}), the new Bianchi identities are
	\begin{equation}
		\tilde{R}^{\mu}_{\cdot [\nu \rho  \sigma ]}=2\tilde{\nabla}_{[\rho} T^{\mu}_{\cdot\nu\sigma ]}-4T^{\lambda}_{\cdot [\nu\rho}T^{\mu}_{\cdot\sigma ]\lambda} \, ,
		\label{bianchi1}
	\end{equation} 
	\begin{equation}
		\tilde{\nabla}_{[\mu |}\tilde{R}^{\alpha}_{\cdot\beta | \nu\rho ]}=-2T^{\lambda}_{\cdot [ \mu \nu |}\tilde{R}^{\alpha}_{\cdot\beta | \rho ] \lambda} \ .
	\end{equation}
	Moreover, it is well known that not all the components of the curvature tensor (\ref{eq:RiemannTilde}) are independent. By definition, this tensor is antisymmetric in the last pair of indices 
	$\tilde{R}^{\mu}_{\cdot \nu \rho  \sigma }=\tilde{R}^{\mu}_{\cdot \nu [\rho  \sigma ]}$. 
	A simple calculation from Eq.~(\ref{RiemannTilde2}) shows that
	\begin{equation}
		\tilde{R}_{(\mu \nu ) \rho  \sigma }=\tilde{\nabla}_{[ \sigma}Q_{\rho ]\mu\nu}+ T^{\lambda}_{\cdot\rho\sigma}Q_{\lambda\mu\nu} \ .
	\end{equation}
	Thus, when the connection is set to be metric-compatible, the curvature tensor is also antisymmetric in the firs pair of indices. 
	The symmetry of the curvature tensor under the exchange of pair of indices depends on the torsion and non-metricity tensors. In general, for non trivial values for those tensors,  this symmetry does not hold. However there are particular conditions under which the exchange symmetry is recovered for non trivial values.

	
	From now on we consider a metric-compatible connection, focusing our attention only on curvature and torsion. We denote by a hat the objects constructed from a metric-compatible connection with torsion:
	\begin{equation}
		\widehat{\Gamma} \equiv \left. \tilde{\Gamma}\right|_{Q=0}.
	\end{equation}
	All the conventions and identities that we have already presented are, of course, still valid.
	The Ricci tensor and the scalar curvature are obtained with the usual contractions, $
	\widehat{R}_{\mu\nu}=\widehat{R}^{\sigma}_{\cdot\mu\sigma\nu}
	$ and $\widehat{R}=g^{\mu\nu}\widehat{R}_{\mu\nu}$. However, the absence of symmetry in the exchange of pair of indices in Eq.~(\ref{eq:RiemannTilde}) allows the Ricci tensor $\widehat{R}_{\mu\nu}$ 
	to be non-symmetric. Indeed, the antisymmetric part of this tensor is
	\begin{equation}
		\widehat{R}_{[ \mu\nu ]}=\widehat{\nabla}_{\rho}( T^{\rho}_{\cdot\mu\nu} +\delta^{\rho}_{\ \mu}T_{\nu}-\delta^{\rho}_{\ \nu}T_{\mu})-2T_{\rho}T^{\rho}_{\cdot\mu\nu}\ .
	\end{equation}
	In view of this identity, a modified torsion tensor can be defined
	\begin{equation}
		\upstar{T}^{\rho}_{\cdot\mu\nu} \equiv  T^{\rho}_{\cdot\mu\nu}+\delta^{\rho}_{\ \mu}T_{\nu}-\delta^{\rho}_{\ \nu}T_{\mu}\, ,
	\end{equation} 
	and a  modified covariant derivative can be introduced,
	\begin{equation}
		\der_{\rho}\equiv \widehat{\nabla}_{\rho} -2T_{\rho} \ .
	\end{equation}
	Hence the antisymmetric part of the Ricci tensor is rewritten as 
	\begin{equation}
		{R}_{[ \mu\nu ]}=\der_{\rho}\upstar{T}^{\rho}_{\cdot\mu\nu} \ .
		\label{eq:AntisymRicci}
	\end{equation}
	It should be stressed the importance of this modified derivative for vectors, since $\partial_{\mu }(\sqrt{-g}A^{\mu})=\sqrt{-g}\,\der_{\mu}A^{\mu}$, for any vector $A^{\mu}$.

	Last but not least, throughout this work we apply the \textit{timelike} convention for the metric signature; i.e. $(+,-,-,-)$.
	

	\section{Quadratic theory of gravity \label{sec:Analisis}}
	
	As we have already argued in the introduction, we are going  to consider an action that is quadratic in the curvature tensor, in order to obtain a theory with propagating torsion \cite{torsionlibro,New,Class,Obukhov,Neville1,Birkov,Flatness}. 
	Excluding parity violating pieces, a total of six independent scalars can be formed from the curvature  tensor (\ref{eq:RiemannTilde}) and its contractions. In addition, other three scalars can be 
	constructed from the torsion tensor (\ref{defTorsion}). On the other hand, the Gauss--Bonnet action is known to lead to a total divergence in a 4-dimensional Riemannian manifold and, therefore, 
	it does not produce any contribution through the variational process of the action. It is worth noting that the Gauss--Bonnet Lagrangian does not contribute to the field equations even in 
	a Riemann--Cartan geometry\footnote{We include the definition of the Gauss--Bonnet action in the presence of torsion and check this property in Appendix \ref{G-B}, since incompatible definitions are used throughout the literature.} \cite{New,GaussBonnet_type}.
	Therefore, the terms  $\widehat{R}^2$, $\widehat{R}_{\nu \sigma} \widehat{R}^{ \sigma \nu}$, and $\widehat{R}_{\mu \nu \rho  \sigma} \widehat{R}^{\rho  \sigma \mu \nu}$ in the Lagrangian density 
	are not independent. Throughout this work, we are going to consider the quadratic Lagrangian density from Poincar\'e gauge theory of gravity, as it is written in Refs.~\cite{New,Class,Birkov,Flatness}. This is

	
	\begin{eqnarray}
		{\cal{L}}_g & =&-\lambda\widehat{R}+\frac{1}{12}(4a+b+3\lambda)T_{\mu \nu \rho}T^{\mu \nu \rho}
		+\frac{1}{6}(-2a+b-3\lambda)T_{\mu \nu \rho}T^{ \nu \rho \mu}
		+\frac{1}{3}(-a+2c-3\lambda)T^{\lambda}_{\cdot\mu \lambda}T_{\rho}^{\cdot\mu \rho }  \nonumber\\
		&+&\frac{1}{6}(2p+q)\widehat{R}_{\mu \nu \rho  \sigma} \widehat{R}^{\mu \nu\rho  \sigma } 
		+\frac{1}{6}(2p+q-6r)\widehat{R}_{\mu \nu \rho  \sigma} \widehat{R}^{\rho  \sigma \mu \nu}
		+\frac{2}{3}(p-q)\widehat{R}_{\mu \nu \rho \sigma}\widehat{R}^{\mu \rho \nu \sigma }\nonumber\\
		&+&(s+t)\widehat{R}_{\nu \sigma} \widehat{R}^{\nu \sigma}
		+(s-t)\widehat{R}_{\nu \sigma} \widehat{R}^{ \sigma \nu} \ ,
		\label{eq:LagrangianoT}
	\end{eqnarray}
	with 
	$\lambda$, $a$, $b$, $c$, $p$, $q$, $r$, $s$ and $t$ the free parameters of the theory. 
	The particular combinations of the parameters that appear in the Lagrangian density have been chosen for convenience without loss of generality. 
	Note that the scalar curvature is also included, which is the only term present in the Einstein--Cartan theory.
	The procedure to obtain the field equations of this Lagrangian density is summarized in Sec.~\ref{sec:ecuaciones}.
	In addition, parity violating pieces can also be assumed in a natural way in the Lagrangian density leading to an interesting results; see Refs.~\cite{Obukhov,PV}. 
	
	In this work we are interested in the stability of theories of gravity with dynamical torsion that avoid large deviations from the predictions of GR where this theory is satisfactory.
	Following this spirit, we focus on quadratic theories, because that is the minimal modification leading to dynamical torsion, and we will not assume that all the components obtained by the irreducible
	decomposition of torsion necessarily propagate. In order to study the stability of the theory, we will focus on two regimes where the metric and torsion degrees of freedom completely decoupled
	from each other through the consideration of the following conditions:
	\begin{enumerate}[a)]
		\item \textit{GR must be recovered when torsion vanishes.}
		\item \textit{The theory must be stable in the weak-gravity regime.}
	\end{enumerate}
	Note that condition $a)$ implies both that the general relativistic predictions will be recovered when torsion is small and that the theory is stable at least when torsion vanishes. 
	This condition will be imposed in Sec.~\ref{sec:reduccionGR} by means of the geometrical structure of the manifold, whereas the second condition will be investigated in Sec.~\ref{sec:stability}
	considering the propagation of torsion modes in a Minkowki space.
	Both conditions have been studied separately in literature using different approaches, see Ref.~\cite{New,Class,Obukhov}.


	\subsection{Field equations\label{sec:ecuaciones}}
	
	The field equations of the Lagrangian density (\ref{eq:LagrangianoT}) have to be obtained, as usual, from a variational principle where the action is extremised with respect to the dynamical variables. 
	However, different sets of dynamical variables can be chosen and different field equations will be obtained accordingly.
	On one hand, the metric and the affine connection can be taken as completely independent variables. Then, the field equations are obtained from varying the action with respect $g^{\mu\nu}$ and $\tilde{\Gamma}^{\sigma}_{\cdot\mu\nu}$. This is called Palatini formalism\footnote{It should be 
		stressed that, for the Palatini method, the general connection $\tilde{\Gamma}$ should be considered. Then, the conditions of metricity and torsion-free must be implemented via Lagrange multipliers.}.
	On the other hand, the connection can be taken to be metric compatible from the beginning. Hence,  the field equations are obtained varying with respect to $g$ and $T$, or 
	to $g$ and $K$. This procedure is sometimes called the metric or Hilbert variational method.
	The Palatini and Hilbert methods are known to differ only on the constraint on the symmetric part of the connection 
	$\tilde{\Gamma}_{(s)}{}^{\sigma}_{\cdot\mu\nu}=\Gamma^{\sigma}_{\cdot\mu\nu}-T^{ \ \mu}_{\nu {}^{.} \sigma}-T^{ \ \mu}_{\sigma {}^{.} \nu}$; that is, they differ on a Lagrange multiplier for the metricity condition, see Refs.~\cite{var1,lagrangemult1}. 
	Therefore, both methods coincide without imposing the Lagrange multiplier when after solving the field equations the related quantity turns to be zero.
	In addition, a third method consists in treating the theory as a gauge theory. This may be seen as being more natural, since the variables are  the gauge potentials $(e_{\mu}^{\ a }, w_{\mu}^{\ a b })$. 
	The field equations in this formalism can been found in Refs.~\cite{Obukhov,Birkov}.

	Let us use the Palatini formalism with the metricity condition implemented as a constraint via a Lagrange multiplier $\Lambda$ to obtain the field equations. 
	The total Lagrangian density of the theory can by written as
	\begin{equation}
		{\cal{L}}={\cal{L}}_g+{\cal{L}}_M+\Lambda^{\ \ \ \rho}_{\nu\mu\cdot}\tilde{\nabla}_{\rho}g^{\mu\nu} \, ,
		\label{LagrangianoTOTAL}
	\end{equation}
	with ${\cal{L}}_g$ from Eq.~(\ref{eq:LagrangianoT}), ${\cal{L}}_M$ the Lagrangian density for matter fields minimally coupled to gravity, and $\Lambda^{\ \ \ \rho}_{\nu\mu\cdot}$ a Lagrange multiplier. 
	The use of the Lagrange multipliers in theories of gravity has been studied in Refs.~\cite{10,8,9}.
	For the sake of simplicity, we rewrite the Lagrangian density ${\cal{L}}_g$  as
	\begin{eqnarray}
		\mathcal{L}_g &=&-\lambda\,\delta_{\alpha}^{\ \gamma}g^{\beta\delta}\tilde{R}^{\alpha}_{\cdot\beta\gamma\delta}
		+f_{{}_{{}_T} \lambda\alpha}^{\ \ \eta\rho\beta\gamma}\ T^{\lambda}_{\cdot\eta\rho}T^{\alpha}_{\cdot\beta\gamma}
		+f_{{}_{{}_R} \lambda\alpha}^{\ \ \eta\rho\sigma\beta\gamma\delta}\ 
		\tilde{R}^{\lambda}_{\cdot\eta\rho\sigma}\tilde{R}^{\alpha}_{\cdot\beta\gamma\delta} \, ,
		\label{DesLgTensPermu}
	\end{eqnarray}
	with the permutation tensors $f_{{}_{{}_T} \lambda\alpha}^{\ \ \eta\rho\beta\gamma}$ and $f_{{}_{{}_R} \lambda\alpha}^{\ \ \eta\rho\sigma\beta\gamma\delta}$ defined in Appendix \ref{variations}. 
	This decomposition factorizes ${\cal{L}}_g$ in parts depending purely on the metric and parts depending on the connection, those are the permutation tensors, and the curvature tensors and the torsion tensors, 
	respectively; thus, the application of Euler-Lagrange equations is straightforward. 
	The field equations for the Lagrangian density (\ref{LagrangianoTOTAL}) are
	\begin{eqnarray}
		\tilde{\mathcal{E}}_{\mu\nu}-(\tilde{\nabla}_{\kappa}-2T_{\kappa})\Lambda_{\nu\mu\cdot}^{\ \ \ \kappa}-\frac{1}{2}\Lambda_{\mu\nu\cdot}^{\ \ \ \kappa}g^{\alpha\beta}\tilde{\nabla}_{\kappa}g_{\alpha\beta}
		&=&\tilde{\tau}_{\mu\nu} \label{eq:LEinstein}, \\
		\tilde{\mathcal{P}}_{\tau}^{\cdot\mu\nu}+2\Lambda_{\tau}^{\cdot\mu\nu}&=&\tilde{\Sigma}_{\tau}^{\cdot\mu\nu}, \label{eq:LPalatini} \\
		\tilde{\nabla}_{\rho}g^{\mu\nu}&=&0 . \label{eq:LMetricidad}
		\label{sistemaEcusLambda}
	\end{eqnarray}
	Note that the metricity condition is obtained as a field equation from the variation of the action with respect to the Lagrange multiplier.
	The definitions used in the above equations are
	\begin{eqnarray}
		\tilde{\mathcal{E}}_{\mu\nu}&\equiv& \frac{1}{\sqrt{-g}}\frac{\partial \sqrt{-g}\mathcal{L}_g}{\partial g^{\mu\nu}}, \\
		\tilde{\mathcal{P}}_{\tau}^{\cdot\mu\nu}&\equiv& \frac{\partial \mathcal{L}_g}{\partial \tilde{\Gamma}^{\tau}_{\cdot\mu\nu}}-
		\frac{1}{\sqrt{-g}}\partial_{\kappa}\left( \sqrt{-g}\frac{\partial \mathcal{L}_g}{\partial  (\partial_{\kappa} \tilde{\Gamma}^{\tau}_{\cdot\mu\nu} ) } \right) \ .
	\end{eqnarray}
	The tensor $\tilde{\mathcal{E}}_{\mu\nu}$ could be considered as the generalization of the Einstein tensor for the Lagrangian density $\mathcal{L}_g$, as it contains the dynamical information of the metric. 
	Analogously, the tensor $\tilde{\mathcal{P}}_{\tau}^{\cdot\mu\nu}$ is the generalization of the Palatini tensor. The source tensors are the  energy-momentum tensor
	\begin{equation}
		\tilde{\tau}_{\mu\nu} \equiv -\frac{1}{\sqrt{-g}} \frac{\partial \sqrt{-g}\mathcal{L}_M (g, \tilde{\Gamma}, \Psi) }{\partial g^{\mu\nu}} \,  ,
		\label{PalatiniSourceG}
	\end{equation}
	and the hypermomentum tensor
	\begin{equation}
		\begin{split}
			\tilde{\Sigma}_{\tau}^{\cdot\mu\nu} \equiv & - \frac{\partial  \mathcal{L}_M (g, \tilde{\Gamma}, \Psi) }{\partial \tilde{\Gamma}_{\cdot\mu\nu}^{\tau}}\, ,
		\end{split}
		\label{PalatiniSourceGamma}
	\end{equation}
	as defined in Refs.~\cite{10,11}.
	
	Now, taking into account the expression of $\mathcal{L}_g$ in Eq.~(\ref{DesLgTensPermu}), the generalized Einstein and Palatini tensors are
	\begin{eqnarray}
		\tilde{\mathcal{E}}_{\mu\nu}&=&-\lambda\tilde{G}_{(\mu\nu )}+\left( \frac{\partial f_{{}_{{}_T} \lambda\alpha}^{\ \ \eta\rho\beta\gamma} }{\partial g^{\mu\nu}} 
		-\frac{1}{2}g_{\mu\nu}f_{{}_{{}_T} \lambda\alpha}^{\ \ \eta\rho\beta\gamma}  \right)T^{\lambda}_{\cdot\eta\rho}T^{\alpha}_{\cdot\beta\gamma} \nonumber\\
		&+&\left( \frac{\partial  f_{{}_{{}_R} \lambda\alpha}^{\ \ \eta\rho\sigma\beta\gamma\delta} }{\partial g^{\mu\nu}} -\frac{1}{2}g_{\mu\nu} f_{{}_{{}_R} \lambda\alpha}^{\ \ \eta\rho\sigma\beta\gamma\delta}  \right)\tilde{R}^{\lambda}_{\cdot\eta\rho\sigma}\tilde{R}^{\alpha}_{\cdot\beta\gamma\delta}  ,
		\label{TildeEconff}
	\end{eqnarray}
	where $\tilde{G}_{(\mu\nu)}$ is the symmetric part of the Einstein tensor, and
	\begin{eqnarray}
		\tilde{\mathcal{P}}_{\tau}^{\cdot\mu\nu}&=&-2\lambda \left[ \upstar{T}^{\nu\mu\cdot}_{\ \ \sigma} +\delta_{\sigma}^{\ \nu}\left( \tilde{\nabla}_{\lambda}g^{\mu\lambda}+
		\frac{1}{2}g^{\alpha\beta}\tilde{\nabla}^{\mu}g_{\alpha\beta}\right)\right.\nonumber\\
		&-&\left.\tilde{\nabla}_{\sigma}g^{\mu\nu}-\frac{1}{2} g^{\mu\nu} g^{\alpha\beta} \tilde{\nabla}_{\sigma} g_{\alpha\beta} \right] \nonumber \\
		&+& 2f_{{}_{{}_T} \lambda\alpha}^{\ \ \eta\rho\beta\gamma} T^{\lambda}_{\cdot\eta\rho}\frac{\partial T^{\alpha}_{\cdot\beta\gamma}}{\partial \tilde{\Gamma}^{\tau}_{\cdot\mu\nu}}
		+2f_{{}_{{}_R} \lambda\alpha}^{\ \ \eta\rho\sigma\beta\gamma\delta} \tilde{R}^{\lambda}_{\cdot\eta\rho\sigma}
		\frac{\partial \tilde{R}^{\alpha}_{\cdot\beta\gamma\delta}}{\partial \tilde{\Gamma}^{\tau}_{\cdot\mu\nu}}\nonumber \\
		&-& \frac{2}{\sqrt{-g}} \partial_{\kappa}
		\left(\sqrt{-g} f_{{}_{{}_R} \lambda\alpha}^{\ \ \eta\rho\sigma\beta\gamma\delta}\tilde{R}^{\lambda}_{\cdot\eta\rho\sigma} \frac{\partial \tilde{R}^{\alpha}_{\cdot\beta\gamma\delta}}{\partial\left(\partial_{\kappa} \tilde{\Gamma}^{\tau}_{\cdot\mu\nu}\right)}  \right),
		\label{TildePconff}
	\end{eqnarray}
	respectively.
	The full expressions of these tensors in terms of the free parameters of the Lagrangian density are shown in Appendix \ref{variations}.
	
	As the metricity condition has arisen as a field equation, from now on we can consider a metric compatible connection $\widehat{\Gamma}$. Then, the field equations (\ref{eq:LEinstein}) and  (\ref{eq:LPalatini}) reduce to
	\begin{eqnarray}
		\widehat{\mathcal{E}}_{\mu\nu}-\der_{\kappa}\Lambda_{\nu\mu\cdot}^{\ \ \ \kappa}&=&\widehat{\tau}_{\mu\nu} \label{eq:LEinsteinM} \\
		\widehat{\mathcal{P}}_{\tau}^{\cdot\mu\nu}+2\Lambda_{\tau}^{\cdot\mu\nu}&=&\widehat{\Sigma}_{\tau}^{\cdot\mu\nu}\ , \label{eq:LPalatiniM} 
	\end{eqnarray}
	To obtain the final expression for the field equations, the Lagrange multiplier $\Lambda$ must be solved out from  Eqs.~(\ref{eq:LEinsteinM}) and (\ref{eq:LPalatiniM}). 
	For this end, note that a generic third rank tensor $A$ can always be written as
	\begin{equation}
		A_{\alpha\beta\gamma}=\Delta_{\beta\alpha\gamma}^{\mu\nu\rho}\left( A_{\mu (\nu\rho )}-A_{[\mu\nu ] \rho}  \right)\  ,
	\end{equation}
	where $\Delta_{\beta\alpha\gamma}^{\mu\nu\rho}$ is defined in Eq.~(\ref{DeltaPermutation}). As $\Lambda^{\ \ \ \rho}_{\nu\mu\cdot}$ is symmetric in the first two indices, we can solve from Eq.~(\ref{eq:LPalatiniM})
	\begin{equation}
		\Lambda_{\mu\nu\rho}=\frac{1}{2}\Delta^{  \alpha\beta\gamma}_{\nu\mu\rho}\left( \widehat{\Sigma}_{\alpha (\beta\gamma )}  -\widehat{\mathcal{P}}_{\alpha (\beta\gamma )}  \right) \ .
	\end{equation}
	Thus, the field equations  become
	\begin{eqnarray}
		\mathcal{\widehat{E}}_{\mu\nu}-\frac{1}{2}\Delta_{\nu\mu\kappa}^{\alpha\beta\gamma}\overline{\nabla}^{\kappa}\left( \widehat{\Sigma}_{\alpha(\beta\gamma )}-\widehat{\mathcal{P}}_{\alpha(\beta\gamma )} \right) &=&\widehat{\tau}_{\mu\nu} \  , \\
		\Delta_{\nu\mu\kappa}^{\alpha\beta\gamma}\left(\widehat{\Sigma}_{[\alpha\beta ]\gamma}-\widehat{\mathcal{P}}_{[\alpha\beta ]\gamma}\right)&=&0 \ .
	\end{eqnarray}
	These are the general expressions of the field equations of any theory of gravity with metricity and torsion. This set of equations is obviously equivalent to the equations obtained from a Hilbert variational principle over the variables $(g, K)$ or $(g, T)$, as can be easily checked.
	Now, taking into account the calculations showed in Appendix \ref{variations} for the Lagrangian density (\ref{eq:LagrangianoT}), these equations are
	
	\begin{eqnarray}
		&-&\lambda\left( \widehat{G}_{(\mu\nu )}-2\overline{\nabla}^{\kappa}\upstar{T}_{(\mu\nu )\kappa} \right) +\frac{1}{12}(4a+b+3\lambda)\left(2T_{\alpha\beta  \mu }T^{\alpha\beta\cdot}_{\ \ \ \nu}-T_{\mu\alpha\beta}T^{\cdot\alpha\beta}_{\nu} -\frac{1}{2}g_{\mu\nu}T_{\alpha\beta\rho}T^{\alpha\beta\rho}\right) \nonumber\\
		&+&\frac{1}{6}(-2a+b-3\lambda)\left(T_{\alpha\beta  \mu }T^{\beta\alpha\cdot}_{\ \ \ \nu} -\frac{1}{2}g_{\mu\nu}T_{\alpha\beta\rho}T^{\beta\rho\alpha} \right)+\frac{1}{3}(-a+2c-3\lambda)\left(T_{\mu}T_{\nu} -\frac{1}{2}g_{\mu\nu}T_{\alpha}T^{\alpha} \right) \nonumber\\
		&+&\frac{1}{6}(2p+q)\left[2\widehat{R}_{\alpha\beta\lambda  \mu} \widehat{R}^{\alpha\beta\lambda \cdot }_{\ \ \ \ \nu}-\frac{1}{2}g_{\mu\nu}\widehat{R}_{\alpha\beta\lambda\sigma} \widehat{R}^{\alpha\beta\lambda\sigma }  -4\overline{\nabla}^{\kappa}\left( \overline{\nabla}^{\lambda}\widehat{R}_{\kappa(\mu\nu )\lambda}+T_{(\mu}^{\cdot \ \lambda\beta}\widehat{R}_{\nu )\kappa\lambda\beta}  \right) \right] \nonumber\\
		&+&\frac{1}{6}(2p+q-6r)\left[ 2\widehat{R}_{\alpha (\mu | \beta\lambda  } \widehat{R}^{\beta\lambda \alpha\cdot }_{\ \ \ \ | \nu )}-\frac{1}{2}g_{\mu\nu}\widehat{R}_{\alpha\beta\lambda\sigma} \widehat{R}^{\lambda\sigma\alpha\beta }  
		-4\overline{\nabla}^{\kappa}\left( \overline{\nabla}^{\lambda}\widehat{R}_{\lambda (\mu\nu )\kappa}+T_{(\mu |}^{\cdot \ \ \lambda\beta}\widehat{R}_{\lambda\beta |\nu )\kappa}    \right)\right] \nonumber\\
		&+&\frac{2}{3}(p-q)\left[ 2\widehat{R}_{\alpha (\mu |\beta\lambda } \widehat{R}^{\alpha\beta \cdot\lambda }_{\ \ \  \ | \nu )}+\widehat{R}_{\alpha\lambda\sigma\mu}\widehat{R}^{\alpha\sigma\lambda\cdot}_{\ \ \ \ \nu} -\widehat{R}_{\mu\alpha\lambda\sigma}\widehat{R}^{\cdot \lambda\alpha\sigma}_{\nu} -\frac{1}{2}g_{\mu\nu}\widehat{R}_{\alpha\beta\lambda\sigma} \widehat{R}^{\alpha\beta\lambda\sigma } \right. \nonumber\\
		&-& \left. 2\overline{\nabla}^{\kappa}\left(   \overline{\nabla}^{\lambda}\widehat{R}_{\kappa (\mu\nu )\lambda}-2T_{\kappa}^{\cdot\lambda\beta}\widehat{R}_{\beta (\mu\nu )\lambda}+2T_{(\mu}^{\ \cdot \lambda\beta}\widehat{R}_{\nu ) \beta\lambda\kappa}-2T_{(\mu |}^{\ \cdot \lambda\beta}\widehat{R}_{\kappa\beta\lambda |\nu )} \right)  \right] \nonumber\\
		&+&(s+t) \left[ \widehat{R}_{\mu\cdot}^{\ \lambda}\widehat{R}_{\nu\lambda}+\widehat{R}^{\lambda}_{\cdot\mu}\widehat{R}_{\lambda\nu }-\frac{1}{2}g_{\mu\nu}\widehat{R}_{\alpha\beta}\widehat{R}^{\alpha\beta}+\overline{\nabla}^{\kappa}\left( g_{\mu\nu}\overline{\nabla}^{\lambda}\widehat{R}_{\kappa\lambda}+ \overline{\nabla}_{\kappa}\widehat{R}_{(\mu\nu )} -  \overline{\nabla}_{(\mu}\widehat{R}_{\nu )\kappa} \right.\right. \nonumber\\
		&-& \left. \left. \overline{\nabla}_{(\mu |}\widehat{R}_{\kappa |\nu )}+\frac{1}{2}T_{(\mu |\kappa\cdot}^{\ \ \ \ \lambda}\widehat{R}_{|\nu ) \lambda}-\frac{1}{2}T_{\kappa(\mu \cdot}^{\ \ \ \ \lambda}\widehat{R}_{\nu ) \lambda}-\frac{1}{2}T_{(\mu\nu )\cdot}^{\ \ \ \ \lambda}\widehat{R}_{\kappa\lambda} \right)  \right] \nonumber\\
		&+&(s-t)\left[    \widehat{R}_{\mu\cdot}^{\ \lambda}\widehat{R}_{\lambda\nu}+\widehat{R}^{\lambda}_{\cdot\mu}\widehat{R}_{\nu\lambda }-\frac{1}{2}g_{\mu\nu}\widehat{R}_{\alpha\beta}\widehat{R}^{\beta\alpha}+\overline{\nabla}^{\kappa}\left( g_{\mu\nu}\overline{\nabla}^{\lambda}\widehat{R}_{ \lambda\kappa} + \overline{\nabla}_{\kappa}\widehat{R}_{(\mu\nu )} -  \overline{\nabla}_{(\mu}\widehat{R}_{\nu )\kappa}\right.\right. \nonumber\\
		&-&\left. \left.  \overline{\nabla}_{(\mu |}\widehat{R}_{\kappa |\nu )} +\frac{1}{2}T_{(\mu |\kappa\cdot}^{\ \ \ \ \lambda}\widehat{R}_{\lambda |\nu ) }-\frac{1}{2}T_{\kappa(\mu | \cdot}^{\ \ \ \ \lambda}\widehat{R}_{ \lambda |\nu )}-\frac{1}{2}T_{(\mu\nu )\cdot}^{\ \ \ \ \lambda}\widehat{R}_{ \lambda\kappa}  \right) \right] = \widehat{\tau}_{\mu\nu}+\frac{1}{2}\Delta_{\nu\mu\kappa}^{\alpha\beta\gamma}\overline{\nabla}^{\kappa}\widehat{\Sigma}_{\alpha (\beta\gamma )} \ ,
		\label{EcFinalG}
	\end{eqnarray}
	and
	\begin{eqnarray}
		&-&2\lambda  \upstar{T}_{\nu\mu\tau}   +\frac{1}{6}(4a+b+3\lambda)T_{[\tau\mu ]\nu}-\frac{1}{6}(-2a+b-3\lambda) \left(   T_{[\mu  \tau]\nu}+T_{\nu\mu  \tau}\right)+\frac{1}{3}(-a+b-3\lambda)   g_{ \nu [\tau}T_{\mu ] } \nonumber\\
		&+&\frac{2}{3}(2p+q)\left( \overline{\nabla}^{\kappa} \widehat{R}_{\tau\mu  \nu\kappa  } -T_{\nu}^{\cdot\lambda\kappa}\widehat{R}_{\tau\mu \lambda\kappa } \right)+\frac{2}{3}(2p+q-6r)\left( \overline{\nabla}^{\kappa} \widehat{R}_{   \nu\kappa  \tau\mu} -T_{\nu}^{\cdot\lambda\kappa}\widehat{R}_{\lambda\kappa\tau\mu} \right) \nonumber\\
		&+&\frac{4}{3}(p-q)\left( \overline{\nabla}^{\kappa} \widehat{R}_{\kappa[\tau\mu ]\nu }-\overline{\nabla}^{\kappa} \widehat{R}_{\nu[\tau\mu ]\kappa }- 2T_{\nu}^{\cdot\lambda\kappa}\widehat{R}_{\kappa[\tau\mu ]\lambda }  \right)   + (s+t) \left( 2g_{\nu [\tau}\overline{\nabla}^{\kappa} \widehat{R}_{\mu ]\kappa}  -2  \der_{[\tau} \widehat{R}_{\mu]\nu}+ T_{\nu\cdot[\tau}^{\ \lambda}\widehat{R}_{\mu ]\lambda}  \right) \nonumber\\
		&+&(s-t) \left(2g_{\nu [\tau |}\overline{\nabla}^{\kappa} \widehat{R}_{\kappa |\mu ]}  -2  \der_{[\tau |} \widehat{R}_{\nu |\mu]}  + T_{\nu\cdot[\tau |}^{\ \lambda}\widehat{R}_{\lambda |\mu ]}   \right)= \widehat{\Sigma}_{[\tau\mu ]\nu}\ .
		\label{EcFinalK}
	\end{eqnarray}
	For an interpretation of the right sides of both field equations see Appendix \ref{Ap:Sources}.

	
	\subsection{Reduction to GR \label{sec:reduccionGR}}
	
	We want to obtain a theory which reduces to GR when torsion vanishes. Thus, the theory will not only be stable in this regime, but it will also deviate only slightly from the predictions of GR
	when torsion is small.
	Note that when torsion is set to zero, the usual Riemannian structure is recovered. Therefore, the Riemann tensor is now symmetric under the exchange of the first and the second pair of indices
	and the Ricci tensor is symmetric. From the  first Bianchi identity (\ref{bianchi1}), it follows
	\begin{equation}
		R_{\mu\nu\rho\sigma}\left(R^{\mu\nu\rho\sigma} -2R^{\mu\rho\nu\sigma} \right)=0 \ \ \textup{for}\ \ T^{\alpha}_{\cdot\beta\gamma}=0 \  .
	\end{equation}
	Then, when $T=0$ the Lagrangian density (\ref{eq:LagrangianoT}) becomes
	\begin{equation}
		\left. {\cal{L}}_g \right|_{T=0}=-\lambda\, R+ (p-r) \, R_{\mu\nu\rho\sigma}R^{\mu\nu\rho\sigma}+2\,s \,R_{\mu\nu } R^{\mu\nu } \ .
	\end{equation}
	From this expression, it is clear that GR is recovered when $T=0$ if and only if $p=r$ and $s=0$. This is the only choice of parameters that leads to GR when torsion vanishes. 
	
	Note that the same conclusion can be extracted from a different and longer approach. That is, considering the field equations (\ref{EcFinalG}) and (\ref{EcFinalK}), it can be concluded that this is 
	the only  choice of parameters that produce the Einstein's equations of GR when torsion vanishes. 
	The same conclusion was achieved in Ref.~\cite{Obukhov}.


	\subsection{Stability in Minkowski spacetime\label{sec:stability}}

	It is well-known that the Lagrangian density (\ref{eq:LagrangianoT}) contains, along with the usual graviton $2^+$, up to six new modes or tordions. 
	These are $2^+$, $2^-$, $1^+$, $1^-$, $0^+$ and $0^-$, in the representation $S^P$ where $S$ is the spin and $P$ is the parity of the mode. A physical meaningful restriction is to demand the theory to be 
	stable in all the $S^P$ sectors, see Refs.~\cite{New,Class,critica1,critica2,Neville2}.
	Quadratic theories in the curvature and torsion tensors are usually treated as a gauge theory, hence the 
	variables considered are the gauge potentials of the Poincar\'e group $(e_{\mu}^{\ a }, w_{\mu}^{\ a b })$. Then, the stability analysis is made through the construction of the spin projection operators.

	In this work, however, we consider the metric formulation.
	We will examine the decoupling limit between the torsion and curvature degrees of freedom.  Thus, in view of Eq.~(\ref{RiemannTilde2}), we focus on the case 
	where $g_{\mu\nu}=\eta_{\mu\nu}$, with $\eta_{\mu\nu}$ the Minkowski metric.  
	For the sake of simplicity, we do not consider the purely tensor component of the torsion in Eq.~(\ref{eq:torsionDESCOMP.}).
	As the only torsion components compatible with a Friedmann-Lema\^itre-Robertson-Walker (FLRW) universe are the vectorial $T^{i}$ and pseudo-vectorial $S^{i}$ components \cite{Gonner:1984rw}, 
	we consider that they are the minimum non-vanishing components that should be taken into account in this framework. In spirit of investigating only slight modifications of GR,
	we assume that they are the only non-vanishing torsion components for a minimal modification over the FLRW background. Under these considerations, we will now impose the absence of ghost and tachyon instabilities for the theory given by 
	the Lagrangian density (\ref{eq:LagrangianoT}). The quadratic Riemann and torsion terms that appear in this Lagrangian density are computed in Appendix \ref{QuadraticTerms}. 
	
	As we consider only the vector and pseudo-vector torsion components in  Minkowski space-time, the Lagrangian density (\ref{eq:LagrangianoT}) reduces in this regime to an ordinary vector and pseudo-vector field theory in  
	flat space-time. A general quadratic action for a vector $A^{\mu}$ in flat space-time comes from 
	\cite{HamiltonianVector,Jimenez:2008sq,Vector theories in cosmology}
	\begin{equation}
		\mathcal{L}=\alpha \partial_{\mu}A_{\nu}\partial^{\mu}A^{\nu}+\beta \partial_{\mu}A_{\nu} \partial^{\nu}A^{\mu} +\gamma \partial_{\mu}A^{\mu}\partial_{\nu}A^{\nu}-\mathcal{V} \ ,
		\label{TeoriaVectorial}
	\end{equation}
	where $\mathcal{V}$ is a possible potential for $A^{\mu}$.
	However, not all the kinetic terms are independent from each other. The terms with factor $\beta$ and $\gamma$ are related by
	\begin{eqnarray}
		\int \sqrt{-g}\,d^4x \ ({\nabla}_{\mu}A^{\mu})^2=
		\int \sqrt{-g}\,d^4x \left( {\nabla}_{\mu}A_{\nu} {\nabla}^{\nu}A^{\mu} 
		+{R}_{\mu\nu}A^{\mu}A^{\nu}\right)\, ,
	\end{eqnarray}
	as can be seen from Eq.~(\ref{eq:IdentRicci}). Thus, in flat space-time these terms are related by a total derivative. 
	On the other hand, as it is well-known, the Hamiltonian density of a system is obtained by performing a Legendre transformation. For this vector system, it is
	\begin{equation}
		\mathcal{H}=\pi^{\mu}\dot{A}_{\mu}-\mathcal{L} \, ,
		\label{legendre}
	\end{equation}
	where $\dot{A}_{\mu}\equiv \partial_0 A_{\mu}$ are the generalized velocities and $\pi^{\mu}$ the canonical momenta defined as $\pi^{\mu}\equiv \frac{\partial \mathcal{L}}{\partial \dot{A}^{\mu}}$. 
	The canonical momenta of the Lagrangian density (\ref{TeoriaVectorial}) are
	\begin{equation}
		\pi^{\mu}=2\alpha \dot{A}^{\mu}+2\beta \eta^{\mu\nu}\partial_{\nu}A^0 +2\gamma \eta^{\mu 0}\partial_{\alpha}A^{\alpha}\, ,
	\end{equation}
	or written through the components of the four-vector,
	\begin{eqnarray}
		\pi^0&=&2(\alpha+\beta+\gamma)\dot{A}^0 +2\gamma \partial_{i}A^{i} \label{pi0a}\, ,\\
		\pi^i&=& 2\alpha\dot{A}^i-2\beta\delta^{ij}\partial_jA^0 \  .
	\end{eqnarray}
	Then, performing the Legendre transformation (\ref{legendre}), the Hamiltonian density reads
	\begin{eqnarray}
		\mathcal{H}&=&\frac{(\pi^0-2\gamma \partial_{i}A^{i})^2}{4(\alpha+\beta+\gamma)}-\frac{(\pi^i+2\beta
			\partial_iA_0)^2}{4\alpha}+ \frac{\beta}{2}F_{ij}F^{ij} \nonumber\\
		&+& \alpha (\partial_iA_0)^2 -(\alpha+\beta)(\partial_iA_j)^2 -\gamma (\partial_iA^i)^2+\mathcal{V} \ ,
	\end{eqnarray}
	with $F_{ij}=2\partial_{[i}A_{j]}$. Unfortunately, the kinetic energy of this system is unbounded from below and, therefore, suffers from ghost-type instabilities whatever the signs of $\alpha$, $\beta$ 
	and $\gamma$ are. This behaviour confirms  that  vector theories suffer from ghost-type instabilities if  all the degrees of freedom  of the four-vector $A^{\mu}$ propagates  (see 
	Refs.~\cite{HamiltonianVector,Jimenez:2008sq}).  Hence, a necessary condition for the absence of this kind of instabilities is to make the scalar mode non-dynamical. Alternatively, the vector degrees of freedom can be 
	frozen and propagate only the scalar mode, but this corresponds to a scalar theory rather than a vectorial one.  To remove the scalar mode, the free parameters of the theory must be chosen in such a way 
	that the canonical momenta given in Eq.~(\ref{pi0a}) vanish. Since $\partial_0{A}^0$ and $\partial_{i}A^{i}$ are independent quantities, the only possibility to cancel out the contribution of $\partial_{i}A^{i}$ to the canonical momenta of the scalar mode is to set $\gamma=0$.
	In addition, $\alpha+\beta=0$ is also needed to remove the contributions of the two remaining kinetic terms in the Lagrangian density (\ref{TeoriaVectorial}) to the dynamics of the scalar mode.
	With these conditions, the  kinetic terms in the vector Lagrangian density becomes a Maxwell-type $ F_{\mu\nu} F^{\mu\nu}$ that only propagates the spatial degrees of freedom of the four-vector $A^{\mu}$. This conclusion is in agreement with the well-known fact that the only ghost-free vector theory in flat space-time is the Maxwell-Proca Lagrangian density. Then, the Hamiltonian density can be positive defined with $\alpha=-\beta<0$. For a more detailed discussion on this item see Ref.~\cite{Vector theories in cosmology}. 
	
	Back to the Lagrangian density (\ref{eq:LagrangianoT}), when the metric corresponds to the Minkowski space-time the expression reduces to 
	\begin{eqnarray}
		{\cal{L}}_g&=&\frac{16}{9} (p+s+t)\partial_{\mu}T_{\nu}\partial^{\mu}T^{\nu}+\frac{16}{9}(p-2r)\partial_{\mu}T_{\nu}\partial^{\nu}T^{\mu} \nonumber\\
		&+&\frac{16}{9}(p-r+5s-t)\partial_{\mu}T^{\mu} \partial_{\nu}T^{\nu} +\frac{1}{9}t\partial_{\mu}S_{\nu}\partial^{\nu}S^{\mu}\nonumber\\
		&-&\frac{1}{9}(2r+t)\partial_{\mu}S_{\nu}\partial^{\mu}S^{\nu} -\frac{1}{18}(3q-4r)\partial_{\mu}S^{\mu}\partial_{\nu}S^{\nu}  \nonumber\\
		&+&\frac{8}{9}(r+t)\varepsilon^{\mu\nu\rho\sigma}\partial_{\rho}T_{\mu}\partial_{\nu}S_{\sigma} -{\cal{V}}(T, S) \, ,
		\label{eq:esta01}
	\end{eqnarray}
	where  ${\cal{V}}(T, S)$ are potential-type terms of the torsion fields, see Appendix \ref{QuadraticTerms}.
	As discussed previously, the free parameters $p$, $q$, $r$, $s$ and $t$ must be carefully selected to produce ghost-free kinetic terms, \textit{ i.e.} Maxwell-type kinetic terms for the trace four-vector $T^{\mu}$ and pseudo-trace four-vector $S^{\mu}$. 
	After suitable integrations by parts the expression above simplifies to
	\begin{eqnarray}
		{\cal{L}}_g&=&  \frac{8}{9}(p+s+t)F_{\mu\nu}(T)F^{\mu\nu}(T) 
		-\frac{1}{18}(2r+t)F_{\mu\nu}(S)F^{\mu\nu}(S) \nonumber\\
		&-&\frac{1}{6}q\partial_{\mu}S^{\mu}\partial_{\nu}S^{\nu}+\frac{16}{3}(p-r+2s)\partial_{\mu}T^{\mu}\partial_{\nu}T^{\nu} -{\cal{V}}(T, S) \  .
		\label{eq:esta1}
	\end{eqnarray}
	Since we have two dynamical fields, there are two canonical momenta. These are
	\begin{eqnarray}
		\pi^{\mu}_T \equiv \frac{\partial \mathcal{L}_g}{\partial (\partial_0 T_{\mu})}&=&\frac{32}{9} (p+s+t)F^{0\mu}(T)
		+\frac{32}{3}\eta^{0\mu}(p-r+2s)\partial_{\alpha}T^{\alpha}\, ,\\
		\pi^{\mu}_S \equiv \frac{\partial \mathcal{L}_g}{\partial (\partial_0 S_{\mu})}&=&-\frac{2}{9}(2r+t)F^{0\mu}(S)-\frac{1}{3} \eta^{0\mu}q\partial_{\alpha}S^{\alpha} \ .
	\end{eqnarray}
	Written through the scalar and vectorial degrees of freedom of the four-vectors
	\begin{eqnarray}
		\pi^{0}_T&=&\frac{32}{3} (p-r+2s)\partial_{\alpha}T^{\alpha}  \ ,\\
		\pi^{i}_T&=&\frac{32}{9}(p+s+t) (\dot{T}^i-\partial^iT^0) \, ,\\
		\pi^{0}_S&=&-\frac{1}{3}q\, \partial_{\alpha}S^{\alpha}  \, ,\\
		\pi^{i}_S&=&-\frac{2}{9}(2r+t) (\dot{S}^i-\partial^iS^0) \  .
		\label{momentos}
	\end{eqnarray}
	As here we have two fields with their own kinetic terms, we need to ensure that neither of them introduces a ghost.
	Thus, to remove the scalar $T^0$ and pseudo-scalar $S^0$ degrees of freedom, we consider $p-r+2s=0$ and $q=0$, respectively. 
	Then, the Hamiltonian density reads
	\begin{eqnarray}
		\mathcal{H}_g& =&-\frac{9}{64}\frac{(\pi_T^i)^2}{(p+s+t)}     -\frac{8}{9}(p+s+t)F_{ij}(T)F^{ij}(T) +\frac{9}{4}\frac{(\pi_S^i)^2}{2r+t} \nonumber\\
		&+&\frac{1}{18}(2r+t)F_{ij}(S)F^{ij}(S)
		+\pi^i_T \partial_i T_o +\pi^i_S \partial_i S_o+{\cal{V}}(T, S) \ .
		\label{HamiltonianoT}
	\end{eqnarray}
	The kinetic energy can be bounded from below with the extra conditions of $p+s+t<0$ and $2r+t>0$  for the vectorial and pseudo-vectorial torsion fields, respectively. These conditions
	are summarized in Table \ref{tab:sol}.

	On the other hand, we now require the absence of tachyon instabilities. 
	In the first place, we consider the weak torsion fields regime, that is the regime where the quadratic terms in torsion fields lead the evolution of the potential.  
	Thus, the potential in the Lagrangian density (\ref{eq:esta1}) takes the form
	\begin{equation}
		{\cal{V}}(T, S)= -\frac{2}{3}(c+3\lambda )T_{\mu}T^{\mu}+\frac{1}{24}( b+3\lambda) S_{\mu}S^{\mu} 
		+{\cal{O}}(3) ,
	\end{equation}
	see Appendix \ref{QuadraticTerms}. Note that the mass terms in an action for a vector field  comes from a potential type 
	$V(\phi)\propto\frac{1}{2}m^2\phi_{\mu}\phi^{\mu}$.  
	Hence, the roles of the squared masses $m^2$ for the vector and pseudo-vector torsion fields are played by the combinations of the 
	coupling constants $b$, $c$ and $\lambda$. For these combinations, the correct sign  must be taken for the spatial components to avoid tachyon-like instabilities. In our convention,  
	$\phi_{\mu}\phi^{\mu}=\phi_0^2-\boldsymbol{\phi}^{2}$, then  $c+3\lambda$ must be positive and $b+3\lambda$ negative for a well-behaved vector and pseudo-vector sector, respectively (see Table \ref{tab:sol}). 
	In summary, with these simple arguments we have found a set of conditions for the ghost and tachyon stability of the Lagrangian density (\ref{eq:LagrangianoT}) at the decoupling limit
	and the weak torsion regime, 
	summarized in Table \ref{tab:sol}.

	In Refs.~\cite{New,Class},  Sezgin and  Nieuwenhuizen provided a detailed analysis of the stability of the Lagrangian density (\ref{eq:LagrangianoT})  for the weak torsion field regime. 
	These two articles were the first systematic stability analysis of this kind of theories, made with the spin projectors formalism, and they are a key reference point in this issue. 
	The conclusions they showed for the stability of both $1^+$ and $1^-$ sectors are compatible with the results obtained in Table \ref{tab:sol}, whereas our analysis to avoid the presence of tachyon instabilities at higher order of the potential provides an additional constraint for the coefficients of the Lagrangian. However, it is worth noting that some authors have suggested that the analysis carried by Sezgin and  Nieuwenhuizen is not restrictive enough to ensure a ghost and tachyon	free spectrum, see Refs.~\cite{critica1,critica2}. In fact, in Ref.~\cite{critica1} the authors pointed out that they even obtain a different expression of the spin projector operator for the 
	pseudo-vector mode. Furthermore, they argue the relevance of considering the additional condition for the absence of $p^{-4}$ poles in all spin sectors, which is not done in the analysis of Refs.~\cite{New,Class}. 
	In Ref.~\cite{fabbri}, Fabbri analyses the stability of the most general quadratic gravitational action with torsion and Dirac fields by demanding, in addition, a consistent decoupling between curvature and torsion that preserves continuity in the torsionless limit, concluding that the only non-vanishing component of torsion is given by the pseudo-vector mode and that  parity-violating terms are not allowed in the Lagrangian density.
	Nevertheless, due to some lack of clarity in the existing literature, a deeper analysis of the origins of these differences is not available yet.
	
	\begin{table*}
		\caption{\small Conditions over the free parameters of the Lagrangian density (\ref{eq:LagrangianoT}) for stability and reduction to GR when torsion vanishes.}
		\label{tab:sol}       
		\begin{tabular}{llll}
			\hline\noalign{\smallskip}
			& $T^{\mu}$ & $S^{\mu}$ & Description  \\
			\noalign{\smallskip}\hline\noalign{\smallskip}
			Ghost-free &  \parbox[c]{2cm}{$p-r+2s=0$\\$p+s+t<0$} &  \parbox[c]{2cm}{$q=0$\\$2r+t>0$} & \parbox[c]{5cm}{To remove the scalar/pseudo-scalar mode and to ensure a well-posed kinetic term.}\\
			\noalign{\smallskip}\noalign{\smallskip}
			Tachyon-free (Weak torsion) &  $c+3\lambda>0$ & $b+3\lambda<0$ & \parbox[c]{5cm}{To have a positive-defined quadratic potential $\mathcal{V}^{(2)}$.} \\
			\noalign{\smallskip}\noalign{\smallskip}
			Tachyon-free (General torsion) & \parbox[c]{2cm}{$p+3s=0$\\$c+3\lambda>0$} & \parbox[c]{2cm}{$p+3s=0$\\$b+3\lambda<0$} &\parbox[c]{5cm}{To cancel $\mathcal{V}^{(4)}$ and to make $\mathcal{V}^{(2)}$ positive-defined. } \\
			\noalign{\smallskip}\hline
			\hline\noalign{\smallskip}
			Reduction to GR when $T^{\alpha}_{\cdot\mu\nu}=0$ & \parbox[c]{2cm}{ $ p-r=0 $ \\$s=0$ } & \parbox[c]{2cm}{ $ p-r=0 $ \\$s=0$ } &  \parbox[c]{5cm}{To recover Einstein-Hilbert action when torsion vaishes. } \\
			\noalign{\smallskip}\hline
		\end{tabular}
	\end{table*}
	\begin{table}
		\caption{\small Compatibility of the constraints for reduction to GR (when $T^\alpha_{\cdot\mu\nu}\to0$) and the stability conditions studied in this paper. In the first column we show necessary conditions for a theory propagating vector or pseudo-vector torsion to be stable.
			Those conditions have to be implemented (at least) by the inequality contained in the second column when the vector mode propagates and by the conditions of the last column when the pseudo-vector also propagates.}
		\label{tab:sol2}       
		\begin{tabular}{lll}
			\hline\noalign{\smallskip}
			Summary & $T^{\mu}$ & $S^{\mu}$  \\
			\noalign{\smallskip}\hline\noalign{\smallskip}
			$p=r=s=0$ &  \parbox[c]{3cm}{$t<0$ \\$c+3\lambda >0$} &  \parbox[c]{3cm}{$q=0$\\$t>0$\\$b+3\lambda <0$} \\
			\noalign{\smallskip}\hline
		\end{tabular}
	\end{table}
	
	Let us now go beyond the weak torsion regime when analysing the potential ${\cal{V}}$. Thus, higher orders in the potential can dominate its evolution.
	The highest order that appears in the potential is quartic, symbolically ${\cal{V}}^{(4)}$, 
	\begin{eqnarray}
		{\cal{V}}^{(4)}(T, S)=&-&\frac{64}{27}(p - r + 2 s) T_{\alpha}T^{\alpha}T_{\beta}T^{\beta}
		-\frac{1}{108}(p - r + 2 s)S_{\alpha}S^{\alpha}S_{\beta}S^{\beta} \nonumber\\
		&+&\frac{8}{81} (2 p + 3 q - 4 r + 2 s)T_{\alpha}S^{\alpha}T_{\beta}S^{\beta} 
		+\frac{8}{81}(p+r+4s)T_{\alpha}T^{\alpha}S_{\beta}S^{\beta} . 
		\label{potencial_4}
	\end{eqnarray} 
	As there are terms mixing the vector and pseudo-vector fields, we note that the potential can be diagonalized in the following basis
	\begin{equation}
		{\cal{V}}^{(4)}=\left( \begin{array}{ccc}
			T_{\alpha}T^{\alpha}  &
			S_{\alpha}S^{\alpha}  &
			T_{\alpha}S^{\alpha}  \end{array}
		\right)
		\mathbb{V}^{(4)}
		\left( \begin{array}{c}
			T_{\alpha}T^{\alpha} \\ S_{\alpha}S^{\alpha} \\ T_{\alpha}S^{\alpha} \end{array} \right),
		\label{potencial_4_1}
	\end{equation}
	with
	$\mathbb{V}^{(4)}$ a $3\times 3$ matrix. The eigenvalues of ${\cal{V}}^{(4)}$ are: 
	\begin{eqnarray}
		\lambda_1&=& -\frac{257}{216}\left( p-r+2s +\sqrt{ A}  \right) , \\
		\lambda_2&=& -\frac{257}{216}\left( p-r+2s -\sqrt{ A}  \right)  ,\\
		\lambda_3&=& \frac{8}{81}(2 p + 3 q - 4 r + 2 s) ,
	\end{eqnarray}
	with
	\begin{eqnarray}
		A&=& \frac{1}{771^2}\left(586249 p^2 - 1168402 p r + 586249 r^2 \right. 
		+ \left. 2349092 p s - 2332708 r s +  2357284 s^2\right) \ .
	\end{eqnarray}
	For a positive-defined quadratic form, the three eigenvalues must be positive. Since we are only interested in the vector and pseudo-vector torsion degrees of freedom, we can assume
	$p-r+2s=0$ and $q=0$, which are the conditions found for making the scalar and pseudo-scalar mode non-dynamic, respectively. Then, the expressions of the eigenvalues reduces to
	\begin{subequations}
		\begin{align}
			& \lambda_1= -\frac{8}{81} (p + 3 s) \, ,\\
			& \lambda_2= \frac{8}{81} (p + 3 s) \, ,\\
			& \lambda_3= -\frac{16}{81} (p + 3 s) \, ,
		\end{align}
	\end{subequations}
	It is easy to see that these eigenvalues cannot be positive at the same time for any combination of $p$ and $s$. Hence, the quartic order in the potential in Eq.~(\ref{HamiltonianoT}) is unstable and,
	therefore, this order must be removed to obtain a stable theory. This can be done taking $3s+p=0$.
	Furthermore, the third order in the potential is not present once we consider that GR is recovered when torsion vanishes. 
	Therefore, when we take $p=r$, $s=0$ and $3s+p=0$, there are only quadratic terms in the potential. 
	Thus, the potential is stable under the same conditions as those obtained in the weak torsion field approximation with the additional constraint of $p+3s=0$, see Table \ref{tab:sol}.
	
	%
	%
	
	On the other hand, we should stress that the stability analysis developed in the literature is usually made using a weak curvature approximation for the metric. However, our stability analysis is made 
	in the limit where the degrees of freedom of the torsion are completely decoupled from those of the metric. For this purpose, we have considered that GR is recovered  when $T=0$  and we have investigated the stability
	of torsion in Minkowski flat space-time, assuming the vector and pseudo-vector modes as the only non-vanishing torsion components. 
	These conditions are combined and summarized in Table \ref{tab:sol2}. 
	Note that when both vector and pseudo-vector modes are simultaneously present, the parameter $t$ must be zero. Consequently, the resulting theory has non-dynamical torsion as only potential-type terms appear in the Lagrangian.

	Finally, we expect that the conditions here obtained, which are found to be necessary and sufficient for the stability of the theory in the decoupling regime, to be necessary but not longer sufficient conditions  when both curvature and torsion are present.
	
	
	\section{Summary \label{sec:conclus}}
	
	In this work we have investigated a quadratic and parity preserving action with curvature and torsion \cite{New,Class,Birkov,Flatness} in order to obtain a stable theory of gravity with dynamical torsion. 
	For this purpose, we have analysed two regimes where the degrees of freedom of the metric and those of the torsion are completely decoupled. The assumptions made in those regimes are also motivated
	by looking for theories whose predictions are expected not to be in great disagreement with those of GR.
	
	On the one hand, we have assumed that the theory reduces to GR when torsion vanishes. This implies the stability of the metric degrees of freedom in the regime where there is no torsion modes.
	Therefore, we have imposed the requirement that the only term independent of the torsion is contained in the scalar curvature $\widehat{R}$, obtaining two conditions for the parameters of the general quadratic	Lagrangian.
	
	On the other hand, we have investigated the stability of the torsion when the metric is flat, following an approach that differs from the usual techniques used in the literature.
	We have focused our attention on the stability of the vector and psuedo-vector torsion components in Minkowski because they are the 
	only components from the torsion irreducible decomposition that propagate in a FLRW spacetime \cite{Gonner:1984rw}. Therefore, it is not necessary to consider the purely tensor component if we are 
	interested in ``minimal'' modifications of the predictions of GR.  
	We have studied the stability of these fields analysing the Hamiltonian formulation of the theory to ensure a ghost and tachyon-free spectrum in this regime. 
	Thus, we have obtained several conditions for the parameters of the general quadratic action (\ref{eq:LagrangianoT}) with propagating torsion that we have summarized in Table \ref{tab:sol}. 
	Furthermore, we have contrasted the conditions obtained in the weak torsion approach with those already presented in the literature \cite{New,Class,critica1,critica2}. As discussed in the previous section, our findings are compatible with the conclusions showed in Refs.~\cite{New,Class}. Finally, we have presented in Table \ref{tab:sol2} the conditions for the stability of the theory combined with the constraints for recovering GR when torsion vanishes.
	
	In summary, we have found the most general subfamily of the Lagrangian density (\ref{eq:LagrangianoT}) that is stable in both decoupling regimes. This is described by
	\begin{eqnarray}
		{\cal{L}}_g & =&-\lambda\widehat{R}+\frac{1}{12}(4a+b+3\lambda)T_{\mu \nu \rho}T^{\mu \nu \rho}
		+\frac{1}{6}(-2a+b-3\lambda)T_{\mu \nu \rho}T^{ \nu \rho \mu}\nonumber\\
		&+&\frac{1}{3}(-a+2c-3\lambda)T^{\lambda}_{\cdot\mu \lambda}T_{\rho}^{\cdot\mu \rho }  ,
		\label{eq:LagrangianoResultadoFinal}
	\end{eqnarray}
	where $b+3\lambda<0$ and $c+3\lambda>0$ and we have restricted ourselves to theories where only the vectorial and pseudo-vectorial torsion components of the irreducible decomposition (\ref{eq:torsionDESCOMP.}) are non zero. However, the resulting theory has a non-dynamical torsion, since no kinetic terms for the torsion are present in the above expression.
	This is because $t$ must be zero in order to simultaneously remove the ghost instabilities from the trace and axial sectors, as their corresponding kinetic terms enter with opposite signs; see Table \ref{tab:sol2}. In consequence, we conclude that the particular case where the quadratic Lagrangian density (\ref{eq:LagrangianoT}) reduces to GR in the absence of torsion cannot safely propagate the vector and pseudo-vector torsion modes simultaneously.


	\section*{Acknowledgement}
	
	The authors acknowledge Jose Beltr\'an Jim\'enez and Francisco Jos\'e Maldonado Torralba for bringing to our attention a missing factor in the pseudo-vectorial sector of the theory in the previous version of the manuscript and for useful discussions.
	The authors also acknowledge Y.~N.~Obukhov for useful discussions.
	This work was partly supported by the projects FIS2014-52837-P (Spanish MINECO) and FIS2016-78859-P (AEI/FEDER, UE), 
	and Consolider-Ingenio MULTIDARK CSD2009-00064.
	PMM was funded by MINECO through the postdoctoral training contract FPDI-2013-16161 during part of this work.

	

	
	\appendix
	\addappheadtotoc
	\appendixpage
	
	\section{The Gauss--Bonnet term in Riemann--Cartan geometries \label{G-B}}
	
	We have noted that there is no agreement about the expression of the Gauss--Bonnet term in a Riemann--Cartan manifold throughout the literature, probably due to several misprints. 
	Therefore, in this appendix, we present the correct expression for the Gauss--Bonnet action. This is:
	\begin{equation}
		S_{GB}=\int d^4x \sqrt{-g}\left(\widehat{R}^2-4\widehat{R}_{\nu \sigma} \widehat{R}^{ \sigma \nu}+ \widehat{R}_{\mu \nu \rho  \sigma} \widehat{R}^{\rho  \sigma \mu \nu}  \right) .
		\label{eq:GB}
	\end{equation}
	
	One can easily check that this is the correct order of the indices focusing attention on the vectorial an pseudo-vectorial torsion fields in the weak curvature approximation. In this regime we have
	\begin{equation}
		\begin{array}{l}
			g_{\mu\nu}=\eta_{\mu\nu}+h_{\mu\nu}  ,\\
			g^{\mu\nu}=\eta^{\mu\nu}-h^{\mu\nu}  .\\
		\end{array}
	\end{equation}
	Let us now prove that, order by order in the fields $h_{\alpha\beta}, T_{\alpha} \ \textup{and} \ S_{\alpha}$, the term (\ref{eq:GB}) leads to a total divergence.
	The expressions of $R^{\mu}_{\cdot\nu \rho  \sigma}$, $R_{\nu\sigma}$ and $R$ in terms of $h$ are well known in linearized gravity  \cite{linearized_h}. These are
	\begin{eqnarray}
		R^{\mu}_{\cdot\nu \rho  \sigma}&=&\frac{1}{2}\left( \partial_{\rho}\partial_{\nu}h^{\mu}_{\ \sigma}+ \partial^{\mu}\partial_{\sigma}h_{\nu\rho}  -\partial^{\rho}\partial^{\mu}h_{\nu\sigma}\right. \nonumber\\
		&-&\left.\partial_{\sigma}\partial_{\nu}h^{\mu}_{\ \rho} \right) ,
		\label{Riemann_h} \\
		R_{\nu\sigma}&=&\frac{1}{2}\left(\partial_{\mu}\partial_{\nu}h^{\mu}_{\ \sigma}  +\partial_{\sigma}\partial_{\mu}h ^{\mu}_{\ \nu}  -\Box h_{\sigma\nu}-\partial_{\sigma}\partial_{\nu}h\right)  ,
		\label{Ricci_h}\\
		R&=&\partial_{\mu}\partial_{\nu}h^{\mu\nu}-\Box h  ,
		\label{Escalar_h}
	\end{eqnarray}
	with $\Box=\partial_{\mu}\partial^{\mu}$.
	Then, from Eq.~(\ref{RiemannTilde2}), it is clear that in the action (\ref{eq:GB}) will appear a Gauss--Bonnet term for the Levi-Civita connection, terms quadratic in torsion and a term mixing torsion and $h$ terms.
	This action can be expressed as
	\begin{eqnarray}
		S_{GB}&=&S_{GB}^{(1)}(\partial h)+S_{GB}^{(2)}(\partial T,\partial S, T, S) 
		+S_{GB}^{(3)}(\partial h,\partial T, T, S)  .
		\label{SGBe}
	\end{eqnarray} 
	The first term on the r.~h.~s.~of this equation is known to be invariant. Nevertheless, this invariance can be proven with an explicit calculation from Eqs.~(\ref{Riemann_h}), (\ref{Ricci_h}) and (\ref{Escalar_h}) 
	with the appropriate boundary conditions on $h$.
	The second term is calculated with the results presented in Appendix \ref{QuadraticTerms}. It can be seen that
	\begin{eqnarray}
		S_{GB}^{(2)}&=&\int d^4x \sqrt{-g}  \left[ \frac{32}{9}(\partial_{\rho}T_{\nu}\partial^{\nu}T^{\rho}-\partial_{\alpha}T^{\alpha} \partial_{\beta}T^{\beta})\right. 
		-\left.\frac{2}{9}(\partial_{\alpha}S^{\alpha} \partial_{\beta}S^{\beta}-\partial_{\alpha}S_{\beta}\partial^{\beta}S^{\alpha})  \right.\nonumber\\
		&+&\frac{64}{27}\partial_{\alpha}\left(T^{\alpha}T_{\beta}T^{\beta}\right)+\left.\frac{4}{27}\partial_{\alpha}\left( T^{\alpha}S_{\beta}S^{\beta} +2 S^{\alpha}T_{\beta}S^{\beta} \right)\right.
		+\left.\frac{8}{9}\epsilon^{	\mu\nu\rho\sigma}\partial_{\nu}S_{\sigma}\partial_{\mu}T_{\rho}\right] .
	\end{eqnarray}
	After integration by parts, the expression above leads to a total divergence. Taking the torsion to be zero at the boundary of ${\cal{U}}_4$, $S_{GB}^{(2)}$ is identically zero. 
	Finally, the third term on the r.~h.~s.~of Eq.~(\ref{SGBe}), $S_{GB}^{(3)}(\partial h,\partial T , T, S)$, is analysed using Eqs.~(\ref{Riemann}), (\ref{Ricci}) and (\ref{EscalarRicci}) for the torsion part and (\ref{Riemann_h}), (\ref{Ricci_h}) and (\ref{Escalar_h}) for the metric dependent part. Thus,
	\begin{eqnarray}
		S_{GB}^{(3)}&=&\int d^4x \sqrt{-g} \left[ 4\Box h \partial_{\alpha}T^{\alpha} -\frac{ }{} 4\partial_{\mu}\partial_{\nu}h^{\mu\nu}\partial_{\alpha}T^{\alpha} 
		\right. \nonumber 
		+\left.\frac{32}{3}\left( \partial_{\sigma}\partial_{\mu}h^{\sigma\mu}\partial_{\alpha}T^{\alpha}-\Box h \partial_{\alpha}T^{\alpha}\right) \right.\nonumber\\
		&+&\left.\frac{8}{3}\left( \partial_{\rho}\partial_{\nu}h\partial^{\rho}T^{\nu}-\partial_{\mu}\partial_{\nu}h^{\mu \sigma}\partial_{\sigma}T^{\nu}\right) \right] .
	\end{eqnarray}
	Note that there are no mixing terms between $\partial h$ and $\partial S$ or $ST$, as it is expected from  parity conservation. After some algebraical manipulations and integration by parts, the equation for $S_{GB}^{(3)}$ vanishes. Hence, we have checked the invariance of an action upon addition of the action (\ref{eq:GB})  in the weak curvature limit. As was pointed by Nieh  \cite{GaussBonnet_type}, 
	the Gauss--Bonnet term will remain invariant even in a curved non-flat metric $g_{\mu\nu}$. But, for  this work, the invariance in weak field limit is sufficient.

	
	\section{Variations in the Palatini formalism \label{variations}}
	
	The Palatini formalism for varying the action consists in taking the metric $g^{\mu\nu}$ and the generic connection $\tilde{\Gamma}^{\sigma}_{\cdot\alpha\beta}$ as the dynamical variables. So, it is useful to rewrite the action in terms of those variables. 
	Some useful well-known relations for considering that variation are 
	\begin{eqnarray}
		g_{\mu\alpha}\delta g^{\alpha\nu} = - g^{\alpha\nu   }\delta g_{\mu\alpha}  ,\qquad \delta \sqrt{-g} = -\frac{1}{2}g_{\mu\nu}\delta g^{\mu\nu}  .
		\label{eq:relations}
	\end{eqnarray}
	Thus, one can easily obtain
	\begin{equation}
		\partial_{\mu}\sqrt{-g}=\frac{1}{2}\sqrt{-g}g^{\alpha\beta}\tilde{\nabla}_{\mu}g_{\alpha\beta}+\sqrt{-g}\tilde{\Gamma}^{\alpha}_{\cdot\alpha\mu}  .
		\label{SinQConT}
	\end{equation} 
	
	Let us know consider the variation of the action written in terms of the Lagrangian density (\ref{DesLgTensPermu}). This is 
	\begin{eqnarray}
		\mathcal{L}_g=&-&\lambda\,\delta_{\alpha}^{\ \gamma}g^{\beta\delta}\tilde{R}^{\alpha}_{\cdot\beta\gamma\delta}
		+f_{{}_{{}_T} \lambda\alpha}^{\ \ \eta\rho\beta\gamma}\ T^{\lambda}_{\cdot\eta\rho}T^{\alpha}_{\cdot\beta\gamma}
		+f_{{}_{{}_R} \lambda\alpha}^{\ \ \eta\rho\sigma\beta\gamma\delta}\ 
		\tilde{R}^{\lambda}_{\cdot\eta\rho\sigma}\tilde{R}^{\alpha}_{\cdot\beta\gamma\delta}.
	\end{eqnarray}
	where the permutation tensors are
	\begin{eqnarray}
		f_{{}_{{}_T} \lambda\alpha}^{\ \ \eta\rho\beta\gamma} &=&  \frac{1}{12}(4a+b+3\lambda)g_{\lambda\alpha}g^{\eta\beta}g^{\rho\gamma}
		+\frac{1}{6}(-2a+b-3\lambda)\delta_{\lambda}^{\ \gamma}\delta_{\alpha}^{\ \eta}g^{\rho\beta}\nonumber\\
		&+&\frac{1}{3}(-a+2c-3\lambda)\delta_{\lambda}^{\ \rho}\delta_{\alpha}^{\ \gamma}g^{\eta\beta}   ,\\
		f_{{}_{{}_R} \lambda\alpha}^{\ \ \eta\rho\sigma\beta\gamma\delta} & =&\frac{1}{6}(2p+q)g_{\lambda\alpha}g^{\eta\beta}g^{\rho\gamma}g^{\sigma\delta} 
		+\frac{1}{6}(2p+q-6r)\delta_{\lambda}^{\ \gamma}\delta_{\alpha}^{\ \rho}g^{\eta\delta}g^{\sigma\beta} 
		+\frac{2}{3}(p-q)g_{\lambda\alpha}g^{\eta\gamma}g^{\rho\beta}g^{\sigma\delta}\nonumber\\ 
		&+&(s+t)\delta_{\lambda}^{\ \rho}\delta_{\alpha}^{\ \gamma}g^{\eta\beta}g^{\sigma\delta}
		+ (s-t)\delta_{\lambda}^{\ \rho}\delta_{\alpha}^{\ \gamma}g^{\eta\delta}g^{\sigma\beta}.
	\end{eqnarray}

	In order to compute the complete generalized Einstein tensor in Eq.~(\ref{TildeEconff}), the following expressions are needed:
	\begin{eqnarray}
		\frac{\partial f_{{}_{{}_R} \lambda\alpha}^{\ \ \eta\rho\sigma\beta\gamma\delta}}{\partial g^{\mu\nu}}&=&\frac{1}{6}(2p+q)\left( \delta_{\mu}^{\ \eta}\delta_{\nu}^{\ \beta}g_{\lambda\alpha}g^{\rho\gamma}g^{\sigma\delta}\right. 
		+\left.\delta_{\mu}^{\ \rho}\delta_{\nu}^{\ \gamma} g_{\lambda\alpha}g^{\eta\beta}g^{\sigma\delta}+\delta_{\mu}^{\ \sigma}\delta_{\nu}^{\ \eta} g_{\lambda\alpha}g^{\eta\beta}g^{\rho\gamma} \right.
		-\left. g_{\alpha\mu}g_{\lambda\nu}g^{\eta\beta}g^{\rho\gamma}g^{\sigma\delta} \right)  \nonumber\\
		&+&\frac{1}{6}(2p+q-6r)
		\left( \delta_{\mu}^{\ \eta}\delta_{\nu}^{\ \delta}\delta_{\lambda}^{\ \gamma}\delta_{\alpha}^{\ \rho}g^{\sigma\beta}\right.
		+\left.\delta_{\mu}^{\ \sigma}\delta_{\nu}^{\ \beta}\delta_{\lambda}^{\ \gamma}\delta_{\alpha}^{\ \rho}g^{\eta\delta}\right)\nonumber\\
		&+&\frac{2}{3}(p-q) 
		\left( -g_{\lambda\mu}g_{\alpha\nu}g^{\eta\gamma}g^{\rho\beta}g^{\sigma\delta}\right.
		+\left.\delta_{\mu}^{\ \eta}\delta_{\nu}^{\ \gamma} g_{\lambda\alpha} g^{\rho\beta}g^{\sigma\delta} 
		+\delta_{\mu}^{\ \rho}\delta_{\nu}^{\ \beta} g_{\lambda\alpha}g^{\nu\gamma} g^{\sigma\delta} \right.
		+\left.\delta_{\mu}^{\ \sigma}\delta_{\nu}^{\ \delta} g_{\lambda\alpha}g^{\nu\gamma} g^{\rho\beta}  \right)\nonumber\\
		&+&(s+t) \left( \delta_{\mu}^{\ \eta}\delta_{\nu}^{\ \beta}\delta_{\lambda}^{\ \rho}\delta_{\alpha}^{\ \gamma}  g^{\sigma\delta}\right.
		+\left. \delta_{\mu}^{\ \sigma}\delta_{\nu}^{\ \delta}\delta_{\lambda}^{\ \rho}\delta_{\alpha}^{\ \gamma}  g^{\eta\beta} \right)\nonumber\\
		&+&(s-t)\left( \delta_{\mu}^{\ \eta}\delta_{\nu}^{\ \delta}\delta_{\lambda}^{\ \rho}\delta_{\alpha}^{\ \gamma}  g^{\sigma\beta} +\delta_{\mu}^{\ \sigma}\delta_{\nu}^{\ \beta}\delta_{\lambda}^{\ \rho}\delta_{\alpha}^{\ \gamma}  g^{\eta\delta} \right) ,\\
		\frac{\partial f_{{}_{{}_T} \lambda\alpha}^{\ \ \eta\rho\beta\gamma} }{\partial g^{\mu\nu}}&=&\frac{1}{12}(4a+b+3\lambda)\left( -g_{\lambda\mu}g_{\alpha\nu} g^{\eta\beta}g^{\rho\gamma}\right. 
		+\left.\delta_{\mu}^{\ \eta}\delta_{\nu}^{\ \beta} g_{\lambda\alpha}g^{\rho\gamma}+\delta_{\mu}^{\ \rho}\delta_{\nu}^{\ \gamma}g_{\lambda\alpha}g^{\eta\beta}  \right)\nonumber\\
		&+& \frac{1}{6}(2p+q-6r)\delta_{\lambda}^{\ \gamma}\delta_{\alpha}^{\ \eta}\delta_{\mu}^{\ \rho}\delta_{\nu}^{\ \beta} 
		+\frac{1}{3}(-a+2c-3\lambda)\delta_{\lambda}^{\ \rho}\delta_{\alpha}^{\ \gamma}\delta_{\mu}^{\ \eta}\delta_{\nu}^{\ \beta} .
	\end{eqnarray}
	For the calculation of the generalized Palatini tensor in Eq.~(\ref{TildePconff}), we need the following expressions:
	\begin{eqnarray}
		\frac{\partial \tilde{R}^{\alpha}_{\cdot\beta\gamma\delta}}{\partial \tilde{\Gamma}^{\tau}_{\cdot\mu\nu}}&=&\tilde{\Gamma}^{\alpha}_{\cdot\tau\gamma}\delta_{\beta}^{\ \mu}\delta_{\delta}^{\ \nu}-\tilde{\Gamma}^{\alpha}_{\cdot\tau\delta}\delta_{\beta}^{\ \mu}\delta_{\gamma}^{\ \nu}+\tilde{\Gamma}^{\mu}_{\cdot\beta\delta}\delta_{\tau}^{\ \alpha}\delta_{\gamma}^{\ \nu}
		-\tilde{\Gamma}^{\mu}_{\cdot\beta\gamma}\delta_{\tau}^{\ \alpha}\delta_{\delta}^{\ \nu}  ,\\
		\frac{\partial T^{\alpha}_{\cdot\beta\gamma}}{\partial \tilde{\Gamma}^{\tau}_{\cdot\mu\nu}}&=&\frac{1}{2}\left(\delta_{\tau}^{\alpha}\delta_{\beta}^{\mu}\delta_{\gamma}^{\nu}- \delta_{\tau}^{\alpha}\delta_{\beta}^{\nu}\delta_{\gamma}^{\mu}  \right)  ,\\
		\frac{\partial \tilde{R}^{\alpha}_{\cdot\beta\gamma\delta}}{\partial\left(\partial_{\kappa} \tilde{\Gamma}^{\tau}_{\cdot\mu\nu}\right)}&=&\delta_{\gamma}^{\kappa}\delta_{\tau}^{\alpha}\delta_{\beta}^{\mu}\delta_{\delta}^{\nu}-\delta_{\delta}^{\kappa}\delta_{\tau}^{\alpha}\delta_{\beta}^{\mu}\delta_{\gamma}^{\nu}  .
	\end{eqnarray}
	Then, taking into account the definition of the torsion and curvature tensors, Eqs.~(\ref{defTorsion}) and (\ref{eq:RiemannTilde}), respectively, the generalized Einstein and Palatini tensors of the quadratic Lagrangian 
	density (\ref{eq:LagrangianoT}) read
	
	\begin{eqnarray}
		\tilde{\mathcal{E}}_{\mu\nu}&=&-\lambda  \tilde{G}_{(\mu\nu )}+\frac{1}{12}(4a+b+3\lambda)\left(2T_{\alpha\beta  \mu }T^{\alpha\beta\cdot}_{\ \ \ \nu}\right.
		-\left.T_{\mu\alpha\beta}T^{\cdot\alpha\beta}_{\nu} -\frac{1}{2}g_{\mu\nu}T_{\alpha\beta\rho}T^{\alpha\beta\rho}\right) \nonumber\\
		&+&\frac{1}{6}(-2a+b-3\lambda)\left(T_{\alpha\beta  \mu }T^{\beta\alpha\cdot}_{\ \ \ \nu} -\frac{1}{2}g_{\mu\nu}T_{\alpha\beta\rho}T^{\beta\rho\alpha} \right)
		+\frac{1}{3}(-a+2c-3\lambda)\left(T_{\mu}T_{\nu} -\frac{1}{2}g_{\mu\nu}T_{\alpha}T^{\alpha} \right)\nonumber\\
		&+&\frac{1}{6}(2p+q)\left(2\tilde{R}_{\alpha\beta\lambda  \mu} \tilde{R}^{\alpha\beta\lambda \cdot }_{\ \ \ \ \nu}-\tilde{R}_{\mu\alpha\lambda\sigma}\tilde{R}^{\cdot\alpha \lambda\sigma}_{\nu}
		+\tilde{R}_{\alpha\mu\lambda\sigma}\tilde{R}^{\alpha \cdot\lambda\sigma}_{\ \nu}-\frac{1}{2}g_{\mu\nu}\tilde{R}_{\alpha\beta\lambda\sigma} \tilde{R}^{\alpha\beta\lambda\sigma }   \right) \nonumber\\
		&+&\frac{1}{6}(2p+q-6r)\left( 2\tilde{R}_{\alpha (\mu | \beta\lambda  } \tilde{R}^{\beta\lambda \alpha\cdot }_{\ \ \ \ | \nu )}
		-\frac{1}{2}g_{\mu\nu}\tilde{R}_{\alpha\beta\lambda\sigma} \tilde{R}^{\lambda\sigma\alpha\beta }  \right) \nonumber\\
		&+&\frac{2}{3}(p-q)\left( 2\tilde{R}_{\alpha (\mu |\beta\lambda } \tilde{R}^{\alpha\beta \cdot\lambda }_{\ \ \  \ | \nu )}+\tilde{R}_{\alpha\lambda\sigma\mu}\tilde{R}^{\alpha\sigma\lambda\cdot}_{\ \ \ \ \nu} 
		-\tilde{R}_{\mu\alpha\lambda\sigma}\tilde{R}^{\cdot \lambda\alpha\sigma}_{\nu} -\frac{1}{2}g_{\mu\nu}\tilde{R}_{\alpha\beta\lambda\sigma} \tilde{R}^{\alpha\lambda\beta\sigma }   \right)  \\
		&+&(s+t) \left( \tilde{R}_{\mu\cdot}^{\ \lambda}\tilde{R}_{\nu\lambda}+\tilde{R}^{\lambda}_{\cdot\mu}\tilde{R}_{\lambda\nu }-\frac{1}{2}g_{\mu\nu}\tilde{R}_{\alpha\beta}\tilde{R}^{\alpha\beta}  \right)
		+(s-t)\left(    \tilde{R}_{\mu\cdot}^{\ \lambda}\tilde{R}_{\lambda\nu}+\tilde{R}^{\lambda}_{\cdot\mu}\tilde{R}_{\nu\lambda }-\frac{1}{2}g_{\mu\nu}\tilde{R}_{\alpha\beta}\tilde{R}^{\beta\alpha} \right) ,\nonumber 
	\end{eqnarray}
	\begin{eqnarray}
		\tilde{\mathcal{P}}_{\tau}^{\cdot\mu\nu}&=& -2\lambda \left[ \upstar{T}^{\nu\mu\cdot}_{\ \ \sigma} +\delta_{\sigma}^{\ \nu}\left( \tilde{\nabla}_{\lambda}g^{\mu\lambda}+\frac{1}{2}g^{\alpha\beta}\tilde{\nabla}^{\mu}g_{\alpha\beta}\right)
		-\tilde{\nabla}_{\sigma}g^{\mu\nu}-\frac{1}{2} g^{\mu\nu} g^{\alpha\beta} \tilde{\nabla}_{\sigma} g_{\alpha\beta} \right]
		+\frac{1}{6}(4a+b+3\lambda)T_{\tau}^{\cdot\mu\nu}\nonumber\\
		&+&\frac{1}{6}(-2a+b-3\lambda) \left(  T_{\ \ \tau}^{\mu\nu\cdot}-T_{\ \ \tau}^{\nu\mu\cdot}  \right)
		+\frac{1}{3}(-a+b-3\lambda)\left(  \delta_{\tau}^{\ \nu}T^{\mu} -\delta_{\tau}^{\ \mu}T^{\nu} \right) \nonumber\\
		&+&\frac{2}{3}(2p+q)\left[ \left(\tilde{\nabla}_{\kappa}-2T_{\kappa} +\frac{1}{2}g^{\alpha\beta}\tilde{\nabla}_{\kappa}g_{\alpha\beta}\right)\tilde{R}_{\tau}^{ \cdot \mu \nu\kappa}
		-T^{\nu}_{\cdot\lambda\kappa}\tilde{R}_{\tau}^{\cdot\mu \lambda\kappa } \right]\nonumber\\
		&+&\frac{2}{3}(2p+q-6r)\left[ \left(\tilde{\nabla}_{\kappa}-2T_{\kappa} +\frac{1}{2}g^{\alpha\beta}\tilde{\nabla}_{\kappa}g_{\alpha\beta}\right)\tilde{R}_{\ \ \ \ \tau}^{ [\nu\kappa ]\cdot\mu}
		-T^{\nu}_{\cdot\lambda\kappa}\tilde{R}_{\ \ \ \ \tau}^{ [\lambda\kappa ]\cdot\mu} \right]\nonumber\\
		&+&\frac{8}{3}(p-q)\left[ \left(\tilde{\nabla}_{\kappa}-2T_{\kappa} +\frac{1}{2}g^{\alpha\beta}\tilde{\nabla}_{\kappa}g_{\alpha\beta}\right)\tilde{R}_{\tau}^{ \cdot   [ \kappa\nu ]\mu}
		-T^{\nu}_{\cdot\lambda\kappa}\tilde{R}_{\tau}^{\cdot[ \kappa\lambda ]\mu}  \right] \\
		&+& (s+t) \left[ 2\delta_{\tau}^{\ \nu}\left(\tilde{\nabla}_{\kappa}- 2T_{\kappa} +\frac{1}{2}g^{\alpha\beta}\tilde{\nabla}_{\kappa}g_{\alpha\beta}\right)\tilde{R}^{\mu\kappa}
		-2\left(\tilde{\nabla}_{\tau}-2T_{\tau} +\frac{1}{2}g^{\alpha\beta}\tilde{\nabla}_{\tau}g_{\alpha\beta}\right)\tilde{R}^{\mu\nu} + T^{\nu}_{\cdot\lambda\tau}\tilde{R}^{\mu\lambda}  \right] \nonumber\\
		&+& (s-t) \left[ 2\delta_{\tau}^{\ \nu}\left(\tilde{\nabla}_{\kappa}-2T_{\kappa} +\frac{1}{2}g^{\alpha\beta}\tilde{\nabla}_{\kappa}g_{\alpha\beta}\right)\tilde{R}^{ \kappa\mu}
		- 2\left(\tilde{\nabla}_{\tau}-2T_{\tau} +\frac{1}{2}g^{\alpha\beta}\tilde{\nabla}_{\tau}g_{\alpha\beta}\right)\tilde{R}^{\nu\mu}  + T^{\nu}_{\cdot\lambda\tau}\tilde{R}^{ \lambda\mu}  \right] \nonumber .
	\end{eqnarray}

	
	\section{Source tensors \label{Ap:Sources}}
	
	In order to understand the right hand side of the field equations, Eqs.~(\ref{EcFinalG}) and (\ref{EcFinalK}), it is necessary to make a distinction between the Hilbert definition of the energy-momentum tensor and the definition carried in Eq.~(\ref{PalatiniSourceG}). 
	The Hilbert's definition is made in a Riemannian $\mathcal{V}_4$ space-time and, therefore, there is a dependence of the matter Lagrangian density on $\partial g$ introduced by the Levi-Civita connection. This definition is
	\begin{eqnarray}
		\tau_{\mu\nu}&\equiv &-\frac{1}{\sqrt{-g}}\frac{\delta\sqrt{-g}{\cal{L}}_M(g, \partial g, \Psi)}{\delta g^{\mu\nu}} 
		=-\frac{1}{\sqrt{-g}}\left(\frac{\partial \sqrt{-g}{\cal{L}}_M}{\partial g^{\mu\nu}} -\partial^{\kappa}\frac{\partial \sqrt{-g}{\cal{L}}_M}{\partial (\partial^{\kappa} g^{\mu\nu})}\right) \ .
		\label{HilbertDef}
	\end{eqnarray}
	Nevertheless, in the Palatini formalism this dependence on $\partial g$ does not exit, since the matter Lagrangian depends  on $g$ and $\tilde{\Gamma}$ as independent variables. Therefore, the energy-momentum tensor reads as in Eq.~(\ref{PalatiniSourceG}). This is
	\begin{equation}
		\tilde{\tau}_{\mu\nu} \equiv -\frac{1}{\sqrt{-g}} \frac{\partial \sqrt{-g}\mathcal{L}_M (g, \tilde{\Gamma}, \Psi) }{\partial g^{\mu\nu}} \, ,
	\end{equation}
	There is a clear difference between both definitions.

	However, when the metricity condition is implemented, the connection $\tilde{\Gamma}$ becomes $\widehat{\Gamma}=\Gamma+K$ and, therefore,  it appears a dependence on  $\partial g$ in the definition (\ref{PalatiniSourceG}). The term $\Delta_{\nu\mu\kappa}^{\alpha\beta\gamma}\overline{\nabla}^{\kappa}\widehat{\Sigma}_{\alpha (\beta\gamma )}$ in the right hand side of Eq.~(\ref{EcFinalG}) takes into account this new dependence that is not present in the original definition of $\widehat{\tau}_{\mu\nu}$.
	To check the consistency of this argument, let us take
	\begin{eqnarray}
		&&\frac{\delta \sqrt{-g}{\cal{L}}_M(g,\partial g, T, \Psi)}{\delta g^{\mu\nu}} =
		\left(\frac{\partial \sqrt{-g}{\cal{L}}_M}{\partial g^{\mu\nu}} -\partial^{\kappa}\frac{\partial \sqrt{-g}{\cal{L}}_M}{\partial (\partial^{\kappa} g^{\mu\nu})}\right) \nonumber\\
		&=&\left(\frac{\partial \sqrt{-g}{\cal{L}}_M}{\partial g^{\mu\nu}} -\partial^{\kappa}\frac{\partial \sqrt{-g}{\cal{L}}_M}{\partial \widehat{\Gamma}_{\alpha}^{\cdot (\beta\gamma )}}\frac{\partial \widehat{\Gamma}_{\alpha}^{\cdot (\beta\gamma )}}{\partial (\partial^{\kappa} g^{\mu\nu})}\right),
	\end{eqnarray}
	where different tensors have been defined in Eqs.~(\ref{levicivita}), (\ref{PalatiniSourceG}) and (\ref{PalatiniSourceGamma}). This leads to
	\begin{equation}
		- \frac{1}{\sqrt{-g}}\frac{\delta\sqrt{-g}{\cal{L}}_M(g,\partial g, T, \Psi)}{\delta g^{\mu\nu}} = 
		\widehat{\tau}_{\mu\nu}+\frac{1}{2}\Delta_{\nu\mu\kappa}^{\alpha\beta\gamma}\overline{\nabla}^{\kappa}\widehat{\Sigma}_{\alpha (\beta\gamma )} \ .
		\label{pruebaTmunu}
	\end{equation}
	The right hand side of Eq.~(\ref{pruebaTmunu}) is exactly the expression on the right hand side of Eq.~(\ref{EcFinalG}), while the left hand side is similar to the Hilbert's definition of the energy-momentum tensor (\ref{HilbertDef}). Indeed $\widehat{\tau}_{\mu\nu}+\frac{1}{2}\Delta_{\nu\mu\kappa}^{\alpha\beta\gamma}\overline{\nabla}^{\kappa}\widehat{\Sigma}_{\alpha (\beta\gamma )}$ is the generalization of the Hilbert's definition of the energy-momentum tensor to the Riemann-Cartan $U_4$ space-time.
	
	On the other hand, $\Sigma_{[\tau\mu]\nu}$ is related to the contortion tensor, which is the remaining part of the connection,  see Ref.~\cite{11}. Thus, the right side of Eq.~(\ref{EcFinalK}) corresponds to the spin distributions tensor
	\begin{equation}
		S_{\sigma}^{\cdot\mu\nu}\equiv -\frac{\partial  \mathcal{L}_M(g, \partial g, T, \Psi)}{\partial K^{\sigma}_{\cdot\mu\nu}} \, ,
	\end{equation}
	as defined in Refs.~\cite{torsionlibro,11}.
	
	
	\section{Vector and pseudo-vector torsion in the weak-gravity regime \label{QuadraticTerms}}
	
	In this appendix we are going to take the vector $T^{\mu}$ and pseudo-vector $S^{\mu}$ torsion components as the only non-vanishing torsion fields and calculate the expressions needed for the analysis carried out in Sec.~\ref{sec:stability}. 
	
	Assuming that the only non-vanishing components of the torsion tensor in the decomposition (\ref{eq:torsionDESCOMP.}) are the vector $T_{\mu}$ and pseudo-vector $S_{\mu}$ torsion componentes, the expression for the contortion tensor (\ref{eq:cotorsion}) can be rewritten as
	\begin{equation}
		K^{\mu}_{{}^{.} \nu \sigma}=\frac{2}{3}g^{\mu \lambda}(T_{\nu}g_{\lambda \sigma}-T_{\lambda}g_{\nu \sigma})+\frac{1}{6}g^{\mu \alpha}\epsilon_{\alpha \nu \sigma \gamma}S^{\gamma} \ .
	\end{equation}
	Under this assumption, the curvature tensor (\ref{RiemannTilde2}) takes the form
	\begin{eqnarray}
		\widehat{R}^{\mu}_{{}^{.} \nu \rho  \sigma} &=&R^{\mu}_{{}^{.} \nu \rho  \sigma}+ \frac{2}{3}\left[\nabla_{\rho}(\delta^{\mu}{}_{\sigma}T_{\nu}-\eta_{\nu \sigma}T^{\mu})\right.
		-\left.\nabla_{\sigma}(\delta^{\mu}{}_{\rho}T_{\nu}-\eta_{\nu \rho}T^{\mu})\right]\nonumber\\
		&+& \frac{4}{9}\left[ (T_{\sigma}T_{\nu}-\eta_{\nu \sigma}T_{\alpha}T^{\alpha})\delta^{\mu}{}_{\rho} \right.
		-\left.(T_{\rho}T_{\nu}-\eta_{\nu \rho}T_{\beta}T^{\beta})\delta^{\mu}{}_{\sigma}+T^{\mu}(T_{\rho}\eta_{\nu \sigma}-T_{\sigma}\eta_{\nu \rho}) \right]\nonumber\\
		&+&\frac{1}{6} \eta^{\mu \alpha}\left(\epsilon_{\alpha \nu \sigma \beta}\nabla_{\rho}S^{\beta} -  \epsilon_{\alpha \nu \rho \beta}\nabla_{\sigma}S^{\beta} \right)
		+\frac{1}{36}\eta^{\mu \alpha}\eta^{\lambda \delta}\left( \epsilon_{\alpha \lambda \rho \tau}\epsilon_{\delta \nu \sigma \gamma}S^{\tau}S^{\gamma} - \epsilon_{\alpha \lambda \sigma \tau}\epsilon_{\delta \nu \rho \gamma}S^{\tau}S^{\gamma} \right) \nonumber\\
		&-&  \frac{1}{9}\left[ T^{\alpha}S^{\gamma}(\delta_{ \ \sigma}^{\mu}\epsilon_{\alpha\nu\rho\gamma}-\delta_{ \ \rho}^{\mu}\epsilon_{\alpha\nu\sigma\gamma}) \right.
		+\left. 2T^{\mu}S^{\gamma}\epsilon_{\rho\nu\sigma\gamma}-2T_{\nu}S^{\gamma}\eta^{\mu\alpha}\epsilon_{\alpha\sigma\rho\gamma}\right.\nonumber\\
		&+&\left.\eta^{\mu\alpha}T^{\lambda}S^{\gamma}(\eta_{\nu\sigma}\epsilon_{\alpha\lambda\rho\gamma}-\eta_{\nu\rho}\epsilon_{\alpha\lambda\sigma\gamma})  \right] \ . 
		\label{Riemann}
	\end{eqnarray}
	The Ricci tensor is obtained by the usual contraction $\widehat{R}^{\mu}_{{}^{.} \nu \mu  \sigma} $, 
	\begin{eqnarray}
		\widehat{R}_{\nu \sigma}  &=&R_{\nu \sigma}-\frac{2}{3}\left(2\nabla_{\sigma}T_{\nu}+\nabla_{\alpha}T^{\alpha}\eta_{\nu \sigma}  \right)
		+\frac{8}{9}\left(T_{\nu}T_{\sigma}-T_{\beta}T^{\beta}\eta_{\nu \sigma}  \right) +\frac{1}{6}\epsilon_{\alpha \nu \sigma \beta}\nabla^{\alpha}S^{\beta} \nonumber\\
		& -& \frac{1}{36}\eta^{\mu \alpha}\eta^{\lambda \delta}\epsilon_{\alpha \lambda \sigma \beta}\epsilon_{\delta \nu \mu \gamma}S^{\beta}S^{\gamma}  ,
		\label{Ricci}
	\end{eqnarray}
	and the scalar curvature  $\widehat{R}=\eta^{\nu \sigma}\widehat{R}_{\nu \sigma} $, 
	\begin{equation}
		\widehat{R} =R-4\nabla_{\alpha}T^{\alpha}-\frac{8}{3}T_{\beta}T^{\beta}+\frac{1}{6}S_{\beta}S^{\beta} \ . 
		\label{EscalarRicci}
	\end{equation}
	
	As we want to get a set of stability condition on the parameters of the theory when $g_{\mu\nu}=\eta_{\mu\nu}$, we take the expression of the curvature tensors (\ref{Riemann}), (\ref{Ricci}) and (\ref{EscalarRicci}) to compute the scalars in the Lagrangian density (\ref{eq:LagrangianoT}). These are

	\begin{eqnarray}
		\left. \widehat{R}^2 \right|_{g=\eta}  &=& 16\partial_{\alpha}T^{\alpha} \partial_{\beta}T^{\beta} +\frac{64}{3}\partial_{\alpha}T^{\alpha}T_{\beta}T^{\beta}
		-\frac{8}{6}\partial_{\alpha}T^{\alpha}S_{\beta}S^{\beta} 
		-\frac{8}{9}T_{\alpha}T^{\alpha}S_{\beta}S^{\beta}\nonumber\\
		&+&\frac{1}{36}S_{\alpha}S^{\alpha}S_{\beta}S^{\beta}+\frac{64}{9}T_{\alpha}T^{\alpha}T_{\beta}T^{\beta} \, , \\
		\left.\widehat{R}_{\nu \sigma} \widehat{R}^{\nu \sigma}  \right|_{g=\eta} &=& \frac{16}{9}\partial_{\mu}T_{\nu}\partial^{\mu}T^{\nu}+\frac{32}{9}\partial_{\alpha}T^{\alpha} \partial_{\beta}T^{\beta}-\frac{1}{18}(\partial_{\alpha}S_{\beta}\partial^{\alpha}S^{\beta}-\partial_{\alpha}S_{\beta}\partial^{\beta}S^{\alpha})-\frac{4}{9}\epsilon^{\mu\nu\rho\sigma}\partial_{\mu}S_{\sigma}\partial_{\rho}T_{\nu}\nonumber\\
		&+&\frac{160}{27}\partial_{\alpha}T^{\alpha}T_{\beta}T^{\beta}-\frac{64}{27}\partial_{\mu}T_{\nu}T^{\mu}T^{\nu} -\frac{10}{27}\partial_{\alpha}T^{\alpha}S_{\beta}S^{\beta}+\frac{4}{27}\partial_{\mu}T_{\nu}S^{\mu}S^{\nu} 
		+\frac{64}{27}T_{\alpha}T^{\alpha}T_{\beta}T^{\beta}\nonumber\\
		&+&\frac{1}{108}S_{\alpha}S^{\alpha}S_{\beta}S^{\beta}- \frac{16}{81}T_{\alpha}T^{\alpha}S_{\beta}S^{\beta}-\frac{8}{81}T_{\alpha}S^{\alpha}T_{\beta}S^{\beta}\, , \\
		\left. \widehat{R}_{\nu \sigma} \widehat{R}^{ \sigma \nu} \right|_{g=\eta}  &= &\frac{48}{9}\partial_{\alpha}T^{\alpha} \partial_{\beta}T^{\beta}+\frac{1}{18}(\partial_{\alpha}S_{\beta}\partial^{\alpha}S^{\beta}-\partial_{\alpha}S_{\beta}\partial^{\beta}S^{\alpha})-\frac{4}{9}\epsilon^{\mu\nu\rho\sigma}\partial_{\mu}S_{\sigma}\partial_{\nu}T_{\rho}+\frac{160}{27}\partial_{\alpha}T^{\alpha}T_{\beta}T^{\beta} \nonumber\\
		&-&\frac{64}{27}\partial_{\alpha}T_{\beta}T^{\beta}T^{\alpha}-\frac{10}{27}\partial_{\alpha}T^{\alpha}S_{\beta}S^{\beta}+\frac{4}{27}\partial_{\mu}T_{\nu}S^{\mu}S^{\nu}+\frac{64}{27}T_{\alpha}T^{\alpha}T_{\beta}T^{\beta}+\frac{1}{108}S_{\alpha}S^{\alpha}S_{\beta}S^{\beta} \nonumber\\
		&-&\frac{16}{81}T_{\alpha}T^{\alpha}S_{\beta}S^{\beta}-\frac{8}{81}T_{\alpha}S^{\alpha}T_{\beta}S^{\beta} \ , \\
		\left. \widehat{R}_{\mu \nu \rho  \sigma} \widehat{R}^{\mu \nu\rho  \sigma } \right|_{g=\eta}  & =&\frac{32}{9}\partial_{\rho}T_{\nu}\partial^{\rho}T^{\nu}+\frac{16}{9}\partial_{\alpha}T^{\alpha} \partial_{\beta}T^{\beta}-\frac{2}{9}\partial_{\alpha}S_{\beta}\partial^{\alpha}S^{\beta}-\frac{1}{9}\partial_{\alpha}S^{\alpha}\partial_{\beta}S^{\beta}+\frac{8}{9}\epsilon^{	\mu\nu\rho\sigma}\partial_{\nu}S_{\sigma}\partial_{\rho}T_{\mu} \nonumber\\
		& -&\frac{128}{27}\partial_{\rho}T_{\nu}T^{\rho}T^{\nu} +\frac{128}{27}\partial_{\alpha}T^{\alpha}T_{\beta}T^{\beta}-\frac{8}{27}\partial_{\alpha}T^{\alpha}S_{\beta}S^{\beta}-\frac{16}{27}\partial_{\alpha}S^{\alpha}T_{\beta}S^{\beta}+\frac{8}{27}\partial_\alpha T_\beta S^\alpha S^\beta   \nonumber\\
		&+&\frac{8}{27}\partial_{\alpha}S_{\beta}T^{\alpha}S^{\beta}+\frac{8}{27}\partial_\alpha S_\beta S^\alpha T^\beta+\frac{64}{27}T_{\alpha}T^{\alpha}T_{\beta}T^{\beta}+\frac{1}{108}S_{\alpha}S^{\alpha}S_{\beta}S^{\beta}-\frac{8}{27} T_{\alpha}T^{\alpha}S_{\beta}S^{\beta} \nonumber\\
		&-&\frac{16}{27}T_{\alpha}S^{\alpha}T_{\beta}S^{\beta}  \, , 
	\end{eqnarray}
	\begin{eqnarray}
		\left. \widehat{R}_{\mu \nu \rho  \sigma} \widehat{R}^{\rho  \sigma \mu \nu} \right|_{g=\eta} & =&\frac{32}{9}\partial_{\rho}T_{\nu}\partial^{\nu}T^{\rho}+\frac{16}{9}\partial_{\alpha}T^{\alpha} \partial_{\beta}T^{\beta}+\frac{2}{9}(\partial_{\alpha}S_{\beta}\partial^{\alpha}S^{\beta}-\partial_{\alpha}S^{\alpha}\partial_{\beta}S^{\beta})+\frac{8}{9}\epsilon^{	\mu\nu\rho\sigma}\partial_{\nu}S_{\sigma}\partial_{\mu}T_{\rho}\nonumber\\
		&-&\frac{128}{27}\partial_{\rho}T_{\nu}T^{\rho}T^{\nu} +\frac{128}{27}\partial_{\alpha}T^{\alpha}T_{\beta}T^{\beta} -\frac{8}{27} \partial_{\alpha}T^{\alpha}S_\beta S^{\beta}+\frac{8}{27}\partial_{\alpha}T_{\beta}S^{\alpha}S^{\beta} \nonumber\\
		&-&\frac{8}{27}\partial_\alpha S_\beta T^\alpha S^\beta -\frac{8}{27}\partial_\alpha S_\beta S^\alpha T^\beta -\frac{8}{27}\partial_\alpha S^\alpha T_\beta S^\beta +\frac{64}{27}T_{\alpha}T^{\alpha}T_{\beta}T^{\beta} \nonumber\\
		&+&\frac{1}{108}S_{\alpha}S^{\alpha}S_{\beta}S^{\beta} +\frac{8}{81}T_{\alpha}T^{\alpha}S_{\beta}S^{\beta}-\frac{32}{81}T_{\alpha}S^{\alpha}T_{\beta}S^{\beta} \, , \\
		\left. \widehat{R}_{\mu \nu \rho \sigma}\widehat{R}^{\mu \rho \nu \sigma } \right|_{g=\eta}  &=&\frac{8}{9}\partial_{\rho}T_{\nu}\partial^{\nu}T^{\rho}+\frac{8}{9}\partial_{\rho}T_{\nu}\partial^{\rho}T^{\nu}+\frac{8}{9}\partial_{\alpha}T^{\alpha} \partial_{\beta}T^{\beta} +\frac{{1}}{6}\partial_{\alpha}S^{\alpha}\partial_{\beta}S^{\beta}\nonumber\\
		&+&\frac{32}{27}\partial_{\alpha}T^{\alpha}T_{\beta}T^{\beta} -\frac{32}{27}\partial_{\alpha}T_{\beta}T^{\beta}T^{\alpha} -\frac{4}{27}\partial_{\alpha}T^{\alpha}S_{\beta}S^{\beta}+\frac{4}{27}\partial_{\alpha}T_{\beta}S^{\alpha}S^{\beta}+\frac{12}{27}\partial_{\alpha}S^{\alpha}T_{\beta}S^{\beta} \nonumber\\
		&+&{\frac{32}{27}}T_{\alpha}T^{\alpha}T_{\beta}T^{\beta}+\frac{1}{216}S_{\alpha}S^{\alpha}S_{\beta}S^{\beta}+\frac{16}{81} T_{\alpha}S^{\alpha}T_{\beta}S^{\beta}\frac{}{}-\frac{4}{81}T_{\alpha}T^{\alpha}S_{\beta}S^{\beta} \ .
	\end{eqnarray}
	Note that there are no terms $\partial T \partial S$, $\partial S TT$, or $STT$ , as it is expected from parity conservation. On the other hand, it is also possible to compute the pure torsion squared terms via Eq.~(\ref{eq:torsionDESCOMP.}). These are,
	\begin{eqnarray}
		T_{\mu\nu\rho}T^{\mu\nu\rho}&=&\frac{2}{3}T_{\mu}T^{\mu}-\frac{1}{6}S_{\nu}S^{\nu},\\
		T_{\mu\nu\rho}T^{\nu\rho\mu}&=&-\frac{1}{3}T_{\mu}T^{\mu}-\frac{1}{6}S_{\nu}S^{\nu},\\
		T^{\lambda}_{\cdot\mu \lambda}T_{\rho}^{\cdot\mu \rho }&=&T_{\mu}T^{\mu} .
	\end{eqnarray}
	
	In view of these calculations, the potential that appears in expression (\ref{eq:esta01}) is
	\begin{eqnarray}
		\mathcal{V}(T, S) &=& -\frac{2}{3} (c+3\lambda)T_{\alpha}T^{\alpha}+\frac{1}{24}(b+3\lambda)S_{\alpha}S^{\alpha}
		+\frac{4}{27}(2p-2q+5s)\partial_\alpha T^\alpha S_\beta S^\beta \nonumber\\
		&-&\frac{8}{27}(p-r+s)\partial_\alpha T_\beta S^\alpha S^\beta+\frac{4}{27}(3q-2r)\partial_\alpha S^\alpha T_\beta S^\beta -\frac{8}{27}r\partial_\alpha S_\beta T^\alpha S^\beta\nonumber\\
		&-&\frac{8}{27}r\partial_\alpha S_\beta  S^\alpha T^\beta 		-\frac{64}{27}(p-r+2s)T_{\alpha}T^{\alpha}T_{\beta}T^{\beta}-\frac{1}{108}(p-r+2s)S_{\alpha}S^{\alpha}S_{\beta}S^{\beta}
		\nonumber\\
		&+&\frac{8}{81}(p+r+4s) T_{\alpha}T^{\alpha}S_{\beta}S^{\beta}+\frac{8}{81}(2p+3q-4r+2s)T_{\alpha}S^{\alpha}T_{\beta}S^{\beta} \ . 
		\label{Potencialtotal}
	\end{eqnarray}
	Note that the parameter $t$ does not appear in the expression of the potential, since the antisymmetric part of the Ricci tensor does not give rise to potential-type terms for the vector and pseudo-vector torsion degrees of freedom.
	

\end{document}